\documentclass{cplslarge}%

\usepackage[authoryear]{natbib}
\usepackage{color}
\usepackage{makeidx}
\makeindex

\begin{document}
\frontmatter
\mainmatter

\author[Marston, Qi \&\ Tobias]{J. B. Marston, Wanming Qi, and S. M. Tobias}
\chapter{Direct Statistical Simulation of a Jet}

\section{INTRODUCTION}
\label{intro}

In this Chapter we review progress that has been made in utilizing one form of Direct Statistical Simulation (DSS) to understand the formation and statistics of jets. We shall first explain the method and place it into context with other statistical procedures, some of which are described in this book. In so doing we shall describe the strengths and weaknesses of the approach when varying degrees of approximation are made. We shall outline some generalizations of DSS and the attendant conservation laws that are preserved for higher-order approximations and describe how these methods compare for the fiducial problem of a stochastically forced jet on a spherical surface. The results will explore the range of validity of quasi-linear approximations for the jet problem, give an insight into the mechanisms that may control jet spacing and strength, and indicate interesting avenues for future research.

Geophysical and astrophysical flows are often far from the theoretician's idealized state of homogeneous isotropic turbulence. The problem of the formation of jets and zonal flows in the oceans, planets and stars (and even the formation of zonal flows in tokamaks) can serve as a testbed for theoretical and computational approaches to the study of such flows. As described elsewhere in this Volume, jets form under conditions of anisotropy --- with jets usually aligning perpendicular to the direction of the variation in rotation --- and inhomogeneity, with jets being a strong function of position.  As the problem of understanding homogeneous and isotropic turbulence is one of the most difficult challenges, it is actually fortuitous that many of the most interesting flows are instead anisotropic and inhomogeneous.  As we discuss below, non-trivial mean flows provide a starting point for a systematic treatment of fluctuations about the flow.  

Any procedure that seeks to explain jets must be able to take into account anisotropy and inhomogeneity; clearly homogeneous and isotropic methods will, by their very nature, lack the ingredients required for a complete description of the interactions that lead to the driving and interactions of jets.  Much progress has been made in our theoretical understanding of these interactions and much of this has been achieved by utilizing experiments (either laboratory or numerical) to elucidate the nonlinear interactions that lead to the complicated dynamics of jet formation. However such approaches are often inefficient for describing the statistical properties of the system; extremely long runs of direct numerical simulations can be required for the convergence of even such simple statistics as the mean behavior. Laboratory experiments have the advantage that they can be run for a long time, but the extraction of information from these experiments is often difficult \citep{readyamazakietal2007}. Thus we follow a different procedure for calculating the statistics of such flows; one for which the statistics are calculated directly rather than {\it a posteriori} from a procedure optimized for calculating the dynamics. We term this procedure Direct Statistical Simulation.

\subsection{A brief introduction to DSS}
\label{introToDSS}

Here we describe a program of research that falls within the framework of DSS as applied to an illustrative problem of a stochastically-driven barotropic jet. We stress that this is a general method, with a range of applicability well beyond this narrow (though certainly important and interesting) application.  In particular it can also be applied to purely deterministic systems \citep{marstonconoveretal2008}.  We shall describe some other areas where DSS may be fruitfully applied in our concluding remarks.  We shall begin by describing the method in very general terms, before specializing to the case of jet formation on a spherical surface for the purpose of illustration.  Our aim is to derive systems that are able to describe successfully the statistical properties of nonlinear systems. We believe that this aim is best served by deriving systematic approximations to the full system that are conservative (i.e. they inherit the conservation laws of the original dynamics) and realizable (with non-negative probability distributions). By systematic we mean simply that the approximations should be an exact representation of the system under certain assumptions (usually an asymptotic limit) and that higher order terms can be introduced in a systematic manner when these assumptions are relaxed and we move away from the asymptotic limit. Finally we stress that, within this framework, it is important to check when DSS does form an adequate replacement for Direct Numerical Simulation (and when the approximations break down) by a careful ``apples to apples'' comparison using the precisely same model. This important step is often lacking from other statistical approaches.

Theories of non-equilibrium statistical mechanics can be broadly grouped into those that attempt to describe correlations at non-equal times, and equal-time formulations.  Non-equal time approaches are generally rather complicated (see \citet{Frisch1995} and the Chapter by Krommes and Parker on ``Statistical Theories'' in this Volume), can be computationally prohibitive \citep{domaradzkietal1987}, and sometimes suffer from divergences \citep{Frisch1995}.  Nevertheless a Quasi-Diagonal Direct Interaction Approximation (QDIA) that simplifies the complicated full DIA has been applied to atmospheric flows \citep{frederiksen1999,okanefrederiksen2004,frederiksen2012b}.  We choose to focus instead on formulations in terms of equal-time statistics.  

Consider a model described by the evolution of a state vector of variables ${\bf q}(\vec{{r}}, t)$ via a partial differential equation (which we term the equation of motion (EOM) for the system). Here we consider the simple case where all the nonlinearities in the system are quadratic and the system is unforced. This system can be written as 
\begin{equation}
{\bf q}_t = {\cal L}[{\bf q}]+{\cal Q}[{\bf q},{\bf q}],
\label{EOM}
\end{equation}
where ${\cal L}$ represents a linear vector differential operator and ${\cal Q}$ is the operator that includes the nonlinear (quadratic) interactions. We proceed by utilizing a standard Reynolds decomposition of the state vector into its mean and fluctuating parts; i.e.\ we set
\begin{equation}
{\bf q} = \overline{{\bf q}}+{{\bf q}^\prime},
\end{equation}
where the average we choose satisfies the Reynolds rules of averaging so that 
\begin{equation}
\overline{\overline{\bf q}}=\overline{\bf q},$  $\overline{{\bf q}^{\prime}} = 0$ and $\overline{\overline{\bf q}~ {\bf q}} = \overline{\bf q}~ \overline{\bf q}.
\end{equation}
Of course many averaging procedures, such as temporal averages and ensemble averages satisfy these rules; in order to simplify matters in this Chapter we restrict our attention to the case where the average is a spatial one over the coordinate along which the jet forms (in this case the zonal direction). Thus the averaging corresponds to a zonal average or a projection onto the zonal wavenumber $m=0$ mode. Once this averaging procedure has been adopted then it is appropriate to talk of mean quantities for the variables with an overbar and fluctuations (or eddies) for the primed variables.

Then on averaging Equation~(\ref{EOM}) we find that
\begin{eqnarray}
\overline{{\bf q}}_t &=& \overline{{\cal L}[{{\bf q}}]}+\overline{{\cal Q}[{\bf q},{\bf q}]}
\\
 &=& {\cal L}[\overline{{\bf q}}]+\overline{{\cal Q}[{\bf q},{\bf q}]},
\label{avEOM}
\end{eqnarray}
where we have used the linearity of ${\cal L}$ and have assumed that the averaging operator commutes with ${\cal L}$.  (It is not always the case that the two operations commute, for example if mass-weighted averaging is employed in a fluid of variable density.  See \citet{aitchaalalmarstonetal2014}.)  Of course the problem is now to describe the average of the nonlinear term in the equation for the mean (which from now on we term the first cumulant). At this point local closure schemes \citep[see e.g.][]{krauseraedler1980,ruediger1988} are often adopted; in their simplest forms these take the form of parameterizing this average as a function of the mean variables \citep[see e.g.][]{canutominotti2001}.  Note however that the presence of derivatives that appear in the quadratic nonlinearity ${\cal Q}$ in Equation~(\ref{avEOM}) means that the two ${\bf q}$ fields should be viewed as at spatially separated points.  A more sophisticated approach is to derive an equation for the evolution of the mean of the fluctuation-fluctuation (or eddy-eddy) interactions.  Again progress is usually made by assuming homogeneity and isotropy of these interactions, whereby analytical expressions can be devised that can be utilized in the equations for the mean quantities. This approach can be extended to the anisotropic case \citep[][]{bartelloholloway1991,maltrudvallis1991}.  We proceed here by making no assumption about the homogeneity or isotropy of the nonlinear interactions, so we include non-local correlations of the form $\overline{{\bf q}'(\vec{r}_1){\bf q}'(\vec{r}_2)}$, where $\vec{r}_1$ and $\vec{r}_2$ are the positions of two vectors within the domain of interest. The equation governing the evolution of this quantity (which we shall term the second cumulant) is obtained by multiplying the EOM  for the system (defined at the point $\vec{r}_1$) by ${\bf q}(\vec{r}_2)$, averaging and symmetrizing. However it is easy to see that this procedure, in addition to introducing quadratic terms from the $\overline{{\bf q}\,{\cal L}[{\bf q]}}$ terms, will require the evaluation of nonlocal cubic terms (which we term the third cumulant) that arise from the $\overline{{\bf q}\,{\cal Q}[{\bf q}, {\bf q}]}$ term in Equation~(\ref{avEOM}). Clearly such terms may be evaluated by extending the procedure to include three-point correlations, i.e.\ multiplying the EOM by ${\bf q}(\vec{{r}_2})$ and ${\bf q}(\vec{{r}_3})$ and averaging and symmetrizing; this leads to evolution equations for the third cumulant that (naturally) includes quartic correlations. If this procedure is repeated it generates an infinite hierarchy of equations describing statistics of higher and higher order. We remind the reader at this point that extending the level of truncation by one order requires the calculation of correlations with an extra point in space, increasing the degrees of freedom of the system by $D$ where $D$ is the dimensionality of the system. It is therefore computationally expedient to truncate this procedure as soon as practical, without oversimplifying the system --- for example by removing conservation laws). Quite how to achieve this is the topic of current research.

The simplest and most computationally efficient approach is to truncate the hierarchy at second order by dropping the contribution of the third cumulant $\overline{{\bf q}'(\vec{r}_1){\bf q}'(\vec{r}_2){\bf q}'(\vec{r}_3)}$ to the tendency of the second cumulant; this truncation of the cumulant expansion at second order is called CE2.  (The term CE2 was introduced in \cite{marston2010}.)  Truncated at this order the equations are realizable, in the sense that energy densities and probabilities are positive and that global conservation laws are respected for quadratic invariants \citep{salmon1998}, because they can be derived as an exact closure of the quasi-linear (QL) approximation (see Section \ref{DNS}). CE2 is quasi-linear in the sense that it includes interactions of mean quantities with eddies to give eddies and interactions of eddies with eddies to give mean flows, but neglects interactions of eddies with eddies to give eddies (which we term eddy-eddy scattering). The set of allowed and forbidden triad interactions is shown in Figure~\ref{triads} and discussed in detail below. This quasi-linear approach allows the evolution of the mean quantities and eddies until a statistical equilibrium has been found. Truncated at this order the equations are formally equivalent to those utilized in Stochastic Structural Stability Theory (SSST or S3T) --- see the Chapters by Farrell and Ioannou, and by Bakas and Ioannou in this Volume as well as \citet{farrellioannou2007,farrellioannou2009,bakasioannou2011,bakasioannou2013,bakasioannou2013b,bakasioannou2013c,bakasioannou2013d,constantinoufarrelletal2014}. 
In addition to supplying the driving energy, within SSST the stochastic forcing is also often used to parametrize contributions from the missing eddy-eddy scattering (in this it differs from CE2).  Also differing from CE2, in SSST changes to the damping are sometimes made to account for the missing eddy -- eddy interactions, an approach taken earlier by \citet{delsole2001}. 

Progress has been made \citep{srinivasanyoung2012,parkerkrommes2013a,parkerkrommes2014} in determining the initial symmetry-breaking bifurcation to jet structures and their subsequent bifurcations (as a pattern forming problem) --- see the Chapter by Parker and Krommes in this Volume.  \citet{bouchetnardinietal2013} have argued that CE2 is an accurate description of the stochastic jet in the limit of large time-scale separation between the fast evolution of the eddies and the slow changes in the mean-flow (see also the Chapter by Bouchet, Nardini and Tangarife in this Volume).  As we shall see below, it is not always the case that the CE2 truncation provides an adequate description of the statistics of the system.  In those cases higher-order terms should be retained. In the simplest of these extensions the evolution equation for the third cumulant should be solved, with the truncation achieved by setting the fourth cumulant equal to zero. This level of truncation is called CE3.  The hierarchy at this order now includes eddy-eddy scattering and therefore allows for the possibility of the description of cascade and inverse-cascade processes. However, the truncation at this order introduces the possibility of lack of realizability of the system, which may be controlled by the introduction of a phenomenological eddy damping parameter or by a projection procedure (both described later). We shall also introduce a computationally less expensive system that includes eddy-eddy scattering which retains only some of the terms in the equations for the third cumulant. We discuss the basic properties of this system which we term CE2.5 in Section \ref{realizability}.  

\section{DSS by Cumulant Expansions}
\label{CE}

\subsection{Coordinate Independent Considerations}
\label{coordinateIndependent}

For concreteness we consider rotating barotropic motion driven at moderate scale and subject to friction $\kappa$ and bi-cubic hyperviscosity with coefficient $\nu_3$.
The EOMs for the low-order cumulants may be written in a
coordinate-independent way.  It is convenient to first rewrite the EOM
in terms of the linear operator $L[A] \equiv  - [\kappa  - \nu_3 (\nabla^2 + 2) 
\nabla^4] A$ and the bilinear Jacobian operator $J[A,~ B] \equiv
\hat{r} \cdot (\vec{\nabla} A \times \vec{\nabla} B)$ as: 
\begin{eqnarray}
\dot{\zeta} = J[\zeta + f,~ \psi] + L[\zeta] + \eta(t),
\label{EOMJ}
\end{eqnarray}
where $\nabla^2 \psi = \zeta$, $f$ is the Coriolis parameter, and $\eta(t)$ is the Gaussian white noise.  The operator $(\nabla^2 + 2)$ that appears in the hyperviscosity ensures that angular momentum 
is conserved on the unit sphere in the absence of friction.  

Direct statistical simulation (DSS) can then be implemented by a Reynolds decomposition of the vorticity into the sum of a mean flow and a fluctuation (a wave or eddy):
\begin{eqnarray}
\zeta(\vec{r}) = \overline{\zeta(\vec{r})} + \zeta^\prime(\vec{r}) \  \ {\rm with}~ \overline{ \zeta^\prime(\vec{r})} = 0 
\label{ReynoldsDecomposition}
\end{eqnarray}
where we choose the averaging operation, denoted by overbar, to be a mean over the longitudinal (zonal) direction.   
The first three equal-time cumulants of the vorticity are given by:
\begin{eqnarray}
c(\vec{r}_1) &\equiv& \overline{\zeta(\vec{r}_1)},
\nonumber \\ 
c(\vec{r}_1, \vec{r}_2) &\equiv&  \overline{\zeta^\prime(\vec{r}_1) \zeta^\prime(\vec{r}_2)}, 
\nonumber \\ 
c(\vec{r}_1, \vec{r}_2, \vec{r}_3) &\equiv&  \overline{\zeta^\prime(\vec{r}_1) \zeta^\prime(\vec{r}_2) \zeta^\prime(\vec{r}_3)}\ .
\label{cumulants}
\end{eqnarray}
We note that the fourth and higher cumulants, unlike the second and third of Equations~(\ref{cumulants}), are not centered moments.  
The fourth cumulant of a Gaussian distribution vanishes, for instance, unlike the fourth centered moment.
The second and higher cumulants contain information about
correlations that are non-local in space, also called ``teleconnection
patterns.''  Perturbative expansion in cumulants is predicated on the
idea that the mean-flow is dominant, and fluctuations about that flow
are small.  Once a closure approximation is
made (see below) the EOM for the cumulants may
be integrated forward in time until a fixed point is reached.
Alternatively the statistics may exhibit slow oscillations that
can then be time-averaged out.

\subsection{Equations of motion for the cumulants}
\label{coordinateIndependentEOM}

Here we investigate DSS at the CE2 and CE3
levels \citep{marstonconoveretal2008,tobiasdagonetal2011} as well as at an intermediate CE2.5 level of approximation.
The EOM for the first cumulant can be obtained most directly by zonal averaging Equation~(\ref{EOMJ}).  This gives an equation of the qualitative form of Equation~(\ref{avEOM}). The second term on the right-hand side of this equation involves the evaluation of the average of the nonlinear Jacobian  $\overline{J[\zeta~, \psi]}$, which
may be re-expressed in terms of the cumulants if the following
auxiliary statistical quantities 
\begin{eqnarray}
p(\vec{r}_1) &\equiv& \overline{\psi(\vec{r}_1)},
\nonumber \\
p(\vec{r}_1, \vec{r}_2) &\equiv& \overline{\zeta^\prime(\vec{r}_1) \psi^\prime(\vec{r}_2)},
\nonumber \\
p(\vec{r}_1, \vec{r}_2, \vec{r}_3) &\equiv& \overline{\zeta^\prime(\vec{r}_1) \zeta^\prime(\vec{r}_2) \psi^\prime(\vec{r}_3)}\ .
\end{eqnarray}
are introduced.  These quantities contain no new information as $c(\vec{r}_1) = \nabla^2_1~ p(\vec{r}_1)$
and $c(\vec{r}_1, \vec{r}_2) = \nabla^2_2~ p(\vec{r}_1, \vec{r}_2)$, where it is understood that the subscript
on differential operators such as $\nabla^2_1$ indicates the variable being differentiated.
Using the identity
\begin{eqnarray}
\psi(\vec{r}_1) = \int \delta(\vec{r}_1 - \vec{r}_2)~  \psi(\vec{r}_2)~ d^2r_2 
\end{eqnarray}
the average Jacobian may be rewritten in such a way that $\psi$ and $\zeta$ are grouped together:
\begin{eqnarray}
\overline{ J_1[\zeta(\vec{r}_1),~ \psi(\vec{r}_1)]} = \int J_1[\overline{ \zeta(\vec{r}_1) \psi(\vec{r}_2) },~  \delta(\vec{r}_1 - \vec{r}_2)]~ d^2r_2\ .
\end{eqnarray} 
Since $\overline{\zeta(\vec{r}_1) \psi(\vec{r}_2)} = p(\vec{r}_1, \vec{r}_2) + c(\vec{r}_1) p(\vec{r}_2)$ the EOM for the first cumulant may then be written in terms of the first and second cumulants as:
\begin{eqnarray}
{{\partial c(\vec{r}_1)}\over{\partial t}} &=& L_1[c(\vec{r}_1)] 
\nonumber \\
&+& \int J_1[p(\vec{r}_1, \vec{r}_2),~ \delta(\vec{r}_1 - \vec{r}_2)] d^2r_2\ .
\label{1stCumulantEOM}
\end{eqnarray}
Here we have used the fact that $J_1[c(\vec{r}_1) + f(\vec{r}_1),~ p(\vec{r}_1)] = 0$ because neither field varies with longitude.  The second term on the right-hand side of Equation~(\ref{1stCumulantEOM}) represents the Reynolds forcing of the mean flow by eddies.  At a fixed point the Reynolds stress is balanced by damping dissipation imparted by the first term on the right-hand side, 
$L_1[c(\vec{r}_1)]$.
The EOM for the second cumulant can be determined by multiplying Equation~(\ref{EOMJ}) by $\zeta(\vec{r}_2)$ followed by zonal averaging.  Closure at the CE2 level, in which the contribution of the third cumulant to the tendency of the second is neglected, yields:
\begin{eqnarray}
{{\partial c(\vec{r}_1, \vec{r}_2)}\over{\partial t}} &=& 2 \bigg{\{} L_1[c(\vec{r}_1, \vec{r}_2)]
\nonumber \\
&+& J_1[c(\vec{r}_1) + f(\vec{r}_1),~ p(\vec{r}_2, \vec{r}_1)] 
\nonumber \\ 
&+& J_1[c(\vec{r}_1, \vec{r}_2),~ p(\vec{r}_1)] \bigg{\}}
+ \Gamma(\vec{r}_1, \vec{r}_2) 
\label{2ndCumulantEOM}
\end{eqnarray}
where $\langle \eta(\vec{r}_1, t_1) \eta(\vec{r}_2, t_2) \rangle = 2 \Gamma(\vec{r}_1, \vec{r}_2)~ \delta(t_1 - t_2)$ is the stochastic covariance matrix with $\langle \ldots \rangle$ denoting time-averaging over a period short compared with the dynamics.  Also  
$\left\{ \right\}$ is short-hand notation for symmetrization that maintains
the invariance of the statistics under interchanges of the field points $c(\vec{r}_2, \vec{r}_1) = c(\vec{r}_1, \vec{r}_2)$; explicitly, 
$\left\{ c(\vec{r}_1, \vec{r}_2) \right\} \equiv \frac{1}{2} [c(\vec{r}_1, \vec{r}_2) + c(\vec{r}_2, \vec{r}_1)]$.   The physical meaning of the terms on the right-hand side of Equation~(\ref{2ndCumulantEOM}) are as follows:  $L_1[c(\vec{r}_1, \vec{r}_2)]$ represents the damping and dissipation of the eddies; the two Jacobians capture the advection of the eddies by the zonal mean flow, and $\Gamma$ is the stochastic driving force upon the eddies.

Going to the next, CE3, level of approximation \citep{marston2012}, we include the contribution of the third cumulant $\overline{\zeta^\prime(\vec{r}_1) \zeta^\prime(\vec{r}_2) \zeta^\prime(\vec{r}_3)}$ to the tendency of the second, and instead impose the requirement that the 4th cumulant vanishes.  The time derivative of the second cumulant, Equation~(\ref{2ndCumulantEOM}), now receives an added contribution from the (now non-zero) third cumulant:
\begin{eqnarray}
{{\partial c(\vec{r}_1, \vec{r}_2)}\over{\partial t}} &=&  \cdots + 2 \bigg\{ \int J_1[p(\vec{r}_1, \vec{r}_2, \vec{r}_3),
\nonumber \\
&&\delta(\vec{r}_1 - \vec{r}_3)]~ d^2r_3 \bigg\}\ .
\label{EOM2from3}
\end{eqnarray}
This contribution captures some effects of eddy -- eddy scattering, and does not affect the conservation of either energy or enstrophy, as discussed below in Section \ref{conservationLaws}.  
For vanishing 4th cumulant, the 4th centered moment equals a sum over products of pairs second cumulants:
\begin{eqnarray}
\overline{\zeta^\prime(\vec{r}_1) \zeta^\prime(\vec{r}_2) \zeta^\prime(\vec{r}_3) \zeta^\prime(\vec{r}_4)} &=& 
c(\vec{r}_1, \vec{r}_2)~ c(\vec{r}_3, \vec{r}_4) 
\nonumber \\
&+& c(\vec{r}_1, \vec{r}_3)~ c(\vec{r}_2, \vec{r}_4)
\nonumber \\
&+& c(\vec{r}_1, \vec{r}_4)~ c(\vec{r}_2, \vec{r}_3)\ .
\end{eqnarray}
As a consequence, the third cumulant now evolves according to:
\begin{eqnarray}
{{\partial c(\vec{r}_1, \vec{r}_2, \vec{r}_3)}\over{\partial t}} &=& 3 \bigg{\{} L_1[c(\vec{r}_1, \vec{r}_2, \vec{r}_3)] 
\nonumber \\
&+ &
J_1[c(\vec{r}_1, \vec{r}_2, \vec{r}_3),~ p(\vec{r}_1)] 
\nonumber \\
&+& J_1[c(\vec{r}_1) + f(\vec{r}_1),~ p(\vec{r}_2, \vec{r}_3, \vec{r}_1)] 
\nonumber \\
&+& 2 J_1[c(\vec{r}_1, \vec{r}_2),~ p(\vec{r}_3, \vec{r}_1)] \bigg{\}}
\nonumber \\
&-& \frac{1}{\tau} c(\vec{r}_1, \vec{r}_2, \vec{r}_3)\ .
\label{3rdCumulantEOM}
\end{eqnarray}
The third cumulant is symmetrized with respect to permutations of the three field points under the $\left\{ \right\}$ operation.  
Parameter $\tau$ is an eddy-damping timescale that models the neglect of fourth cumulant \citep{orszag1977}.  Here we consider only the simplest case of constant eddy-damping rate; more complicated choices are also possible.  In the literature on homogeneous and isotropic turbulence, these choices are typically guided by a desire to reproduce power-law scaling of the power spectrum; see for instance \citet{davidson2004}.  For damped and driven flows damping must be included as otherwise the time-evolved CE3 equations blow up.  Note that CE2 is recovered in the $\tau \rightarrow 0$ limit that suppresses the third cumulant so CE3 can be controlled at small $\tau$.   We also note the recent investigation of an alternative higher-order closure in the context of uncertainty quantification \citep{SapsisMajda2013a,SapsisMajda2013b}.

\subsection{Realizability}
\label{realizability}

Conservation of energy and enstrophy in the absence of forcing and dissipation are enough to guarantee stability \citep{arakawa1966} in the sense that runaway behavior does not occur.  However conservation laws alone are not sufficient to ensure that the probability distribution function for the fields remains non-negative.  Closures are generally non-realizable:  They cannot be exactly realized by an auxiliary linear model \citep{salmon1998} and therefore can and often do develop negative probability densities.   When forcing and dissipation are present, runaways can occur as these external reservoirs of energy may feed into the negative probability modes.  To prevent such pathologies closure at the CE3 level must be modified.  We present two alternatives that have complementary strengths and weaknesses.  

The first approach, denoted\footnote{We thank Bill Young for suggesting this terminology.} CE3$^*$, is based upon the fact that the eigenvalues $\lambda_i$ of the second cumulant cannot be negative \citep{kraichnan1980}.  It is easy to see that this must be the case by working in a basis $\varphi_i(\vec{r})$ in which the second cumulant, which is self-adjoint, is diagonal.  As the diagonal entries are averages of squares, each must be non-negative for realizable (non-negative) probability distributions.  In CE3$^*$ the equations of motion for the first through third cumulants are integrated forward in time, and at regular intervals all eigenvectors with negative eigenvalues are excised from the second cumulant:
\begin{eqnarray}
c(\vec{r}_1, \vec{r}_2) &=& \sum_i \lambda_i~ \varphi_i(\vec{r}_1)~ \varphi_i(\vec{r}_2) 
\nonumber \\
&\rightarrow& \approx \sum_{i,~ \lambda_i > 0} \lambda_i~~ \varphi_i(\vec{r}_1) \varphi_i(\vec{r}_2)\ .
\label{projection}
\end{eqnarray}
For the simulation reported here the projection is carried out every 2 time steps, but the statistics are insensitive to the interval provided that it is short enough to ensure stability.
The approach has the advantage that it introduces no phenomenological parameters (eddy damping can be eliminated altogether, or $\tau$ can be kept finite if tuning is desired), but it comes at the cost of violating the conservation of energy and enstrophy in the limit of no forcing or dissipation.  Nevertheless, the violations can be small, and as we show below, sensible results can be obtained.  It may be interesting to explore whether realizability can instead be imposed on the third cumulant \citep{kraichnan1980} as that would preserve the conservation of the quadratic invariants (see Section \ref{conservationLaws} below).

An alternate approach is based upon an approximation intermediate between CE2 and CE3 that we call CE2.5.  It can be obtained by setting the left-hand side of Equation~(\ref{3rdCumulantEOM}) to zero,  equivalent to the assumption that the third cumulant adjusts quickly enough to remain near equilibrium while the second and first cumulants vary more slowly.  (We are currently checking under what circumstances this assumption holds in direct numerical simulation of jet formation on the $\beta$-plane.) Thus the prognostic equation for the third cumulant is replaced by a diagnostic equation, similar to the eddy-damped quasi-normal Markovian (EDQNM) approximation \citep{orszag1970}. Upon further neglecting the first three terms on the right-hand side of Equation~(\ref{3rdCumulantEOM}), which represent the linear terms renormalized 
by the mean flow\footnote{This approximation is done for computational expediency, to avoid solving a non-trivial linear equation.  It would be interesting to investigate retaining the linear terms.}, a closed expression for the third cumulant in terms of the second is obtained:
\begin{eqnarray}
c(\vec{r}_1, \vec{r}_2, \vec{r}_3) = 6 \tau \left\{ J_1[c(\vec{r}_1, \vec{r}_2),~ p(\vec{r}_3, \vec{r}_1)] \right\}
\label{CE2.5}
\end{eqnarray}
Equation~(\ref{CE2.5}) may then be substituted into Equation~(\ref{EOM2from3}), eliminating the third cumulant altogether and yielding another closure at the level of second cumulants, yet one that (unlike CE2) includes eddy-eddy scattering.  Furthermore Equation~(\ref{EOM2from3}) shows that the added contribution to the tendency of the second cumulant is similar to that of the stochastic forcing.  This is the ``Markovian'' part of EDQNM, which has been generalized to anisotropic (but still homogeneous) flows \citep{legras1980,bowmankrommesetal1993,herretal1996,bowmankrommes1997,bertoglio2003}.  It appears (but is not proven) that CE2.5 provides the generalization of EDQNM to inhomogeneous flows.  Numerical experiments (such as the ones discussed below) find that the second cumulant does not develop negative eigenvalues for $\tau$ not too large.  It would be good to know if that can be demonstrated rigorously.  

As discussed in the next subsection, CE2.5, like CE3, conserves angular momentum, energy, and enstrophy and is thus better than CE3$^*$ in this regard.  On the other hand by construction CE2.5 includes at a minimum one phenomenological parameter ($\tau$) that must be tuned, and thus has less predictive power than CE3$^*$.    

\subsection{Conservation Laws Respected by CE Closures}
\label{conservationLaws}
In the absence of forcing and dissipation, the 2D barotropic incompressible fluid conserves global angular momentum, energy and an infinite hierarchy of Casimirs.  It is thus of theoretical interest to understand whether conservation laws of the exact dynamics are preserved by the different CE closures. It is straightforward to show \citep{legras1980} that both energy and enstrophy are conserved by each set of three triads that describe scattering between the same three wavevectors. 
Decimation of sets of triad interactions changes the distribution of energy and enstrophy among wavenumbers, but maintains overall conservation of the global quantities.
Since the CE2 assumption amounts to the neglect of eddy-eddy interactions and thus retention of only a subset of all the sets of triads, it must also conserve total energy and total enstrophy.  This argument does not directly apply to CE2.5 or CE3, or to the higher Casimirs, and a closer look at the conservation laws is required.  

Consider 2D barotropic incompressible inviscid fluid living on a general boundaryless surface embedded in the three-dimensional (3D) Euclidean space. The 2D surface is defined as $n(x,y,z)=1$ in terms of the 3D Cartesian coordinates ${\vec{r}} = (x,y,z)$, where $n$ is a general coordinate and $\hat{n}$ is normal to the surface. The time evolution of the relative vorticity field $\zeta$ on the 2D surface is governed by the EOM
\begin{eqnarray}
\frac{\partial \zeta}{\partial t} = J[\zeta+ f, \psi],
\label{ConservativeEOM}
\end{eqnarray} 
where relative vorticity field $\zeta$ is related to the relative streamfunction $\psi$ through $\zeta = \nabla^2 \psi$, $f$ is taken here as a general function on the 2D surface, and the Jacobian operator is again $J[A, B] = \hat{n}\cdot (\vec{\nabla} A \times \vec{\nabla} B) $.  The EOM shows that the quantity (absolute vorticity) $q = \zeta +f$ is transported by a velocity field ${\bf{u}} =  \hat{n} \times \vec{\nabla} \psi$ along the iso-streamfunction lines.

Choose appropriate coordinates $(\mu,\nu)$ on the 2D surface so that the unit vectors $(\hat{\mu},\hat{\nu},\hat{n})$ form a locally orthogonal right-handed basis. 
 The appropriate curvilinear coordinates $(\mu,\nu,n)$ should further satisfy the following five conditions:
\begin{enumerate}
   \item The product of the scale factors on the surface $h_\mu (\mu,\nu, n=1) \cdot h_\nu (\mu,\nu,n=1) $ is independent of the coordinate $\mu$, that is, $h_\mu (\mu,\nu, n=1) \cdot h_\nu (\mu,\nu,n=1) = h(\nu)$, where $h$ is a function of $\nu$. Here $h_\mu$ and $h_\nu$ are scale factors of the coordinates $\mu$ and $\nu$ respectively: $h_\mu \equiv |\partial {\vec{r}}/\partial \mu|$ and $h_\nu \equiv |\partial {\vec{r}}/\partial \nu|$. Then the $\mu$-direction is denoted as the zonal direction and the geometry has zonal symmetry.
   \item The function $f$ has the form $f(\mu,\nu) = \beta \nu$, where $\beta$ is a constant. Then the dynamics also has zonal symmetry. 
   \item The range of $\mu$ on the surface is $[0,L_\mu]$, where $L_\mu$ is a constant independent of $\nu$. Otherwise, the definition of the zonal averages for many-point correlations is problematic as the $\mu$ ranges for different positions can be different.
   \item The range of $\nu$ on the surface is $[\nu_a, \nu_b]$, where $\nu_a$ and $\nu_b$ are constants independent of $\mu$. Otherwise the global invariants of the form $\int d^2r F(\mu,\nu)$ cannot be expressed in terms of the zonal averages by integrating over $\mu$ first.
  \item The boundaryless surface is either sphere-like or torus-like. The sphere-like boundary condition is defined as periodic boundary condition for $\mu$, and all $\mu$ shrink to one point at the boundary of $\nu$. The torus-like boundary condition means periodic boundary conditions for both $\mu$ and $\nu$.
\end{enumerate}
Note that this formulation includes both the case of the sphere for which $(\mu,~ \nu,~ n) = (\phi,~ \cos\theta,~ r)$, and the doubly-periodic $\beta$-plane where $(\mu,~ \nu,~ n) = (x,~ y,~ z)$. For surfaces that satisfy the above-mentioned conditions, the EOMs of zonally-averaged cumulants at CE2, CE2.5 or CE3 (no CE3$^*$ projection) levels of approximation conserve one or two linear quantities, namely the circulation: 
\begin{eqnarray}
\Gamma_1 \equiv \int d^2r~ \zeta(\mu,\nu)
\end{eqnarray}
and on the sphere the $z$-component of the angular momentum:
\begin{eqnarray}
L_z \equiv \int d^2r~ \zeta(\mu,\nu)~ \nu\ .
\end{eqnarray}
Two quadratic quantities, the total (kinetic) energy 
\begin{eqnarray}
E \equiv - \frac{1}{2} \int  d^2r~\zeta(\mu,\nu)~\psi(\mu,\nu)~,
\end{eqnarray}
and the absolute enstrophy
\begin{eqnarray}
\Gamma_2 \equiv \int d^2r~ q^2 (\mu,\nu)
\end{eqnarray}
are also conserved.  
Moreover, CE3 with no eddy-damping ($\tau \rightarrow \infty$) further respects the conservation of the third Casimir of absolute vorticity
\begin{eqnarray}
\Gamma_3 \equiv \int d^2r~ q^3 (\mu,\nu).
\end{eqnarray}
as can be shown by detailed calculation \citep{qi2014}.

The result can be generalized to any CE-N closure that neglects the $(N+1)$-th and higher-order cumulants.  Any CE-N ($N \geq 2$) not only conserves the linear and quadratic invariants, but also conserves up to the $N$-th Casimir.  The proof of this is technical, but we give a brief sketch here; details can be found in \citet{qi2014}.  It is easy to understand that
CE-N conserves up through and including the $(N-1)$-th Casimir; for example, a quadratic invariant such as enstrophy can be expressed in terms of the first and second cumulants, so its time derivative only depends on the EOMs for the first and second cumulants. Note that CE3 and higher-order closures do not affect the first and second cumulant EOMs in the hierarchy; the two EOMs are the same as those in the infinite hierarchy.  We can infer by this kind of reasoning that CE-N at least conserves up to $(N-1)$-th Casimir.  A non-trivial calculation is then required to show that the modification of the N-th cumulant EOM by CE-N closure assumption does not affect the conservation of the $N$-th Casimir \citep{qi2014}.  

The higher Casimirs play an important role in the equilibrium statistical mechanics of inviscid 2D turbulence (see \cite{qimarston2014} and references therein).  It may be interesting to investigate how that physics relates to the non-equilibrium statistical mechanics of jets.  But the conservation can be put to immediate practical use by testing the correctness of the numerical implementation of the cumulant expansions.  In the conservative limit of no forcing and dissipation, the conservation of circulation, angular momentum, energy, and enstrophy tightly constrains CE2, and coding or other errors immediately become evident as violations of the conservation laws.  CE3 must additionally respect conservation of the third Casimir, providing a stringent test of its implementation.  Unlike the linear and quadratic invariants, however, finite resolution (in either real or spectral space) breaks conservation of $\Gamma_3$.  To use the invariance, we integrate the DNS and CE3 equations of motion starting from an initial condition that has power at only low wavevectors.  Then for short times the third Casimir is conserved, and errors in the equation of motion for DNS and CE3 can be readily detected.

\subsection{Basis of Spherical Harmonics} 
\label{sphericalHarmonics}

\begin{figure}%
\figurebox{20pc}{}{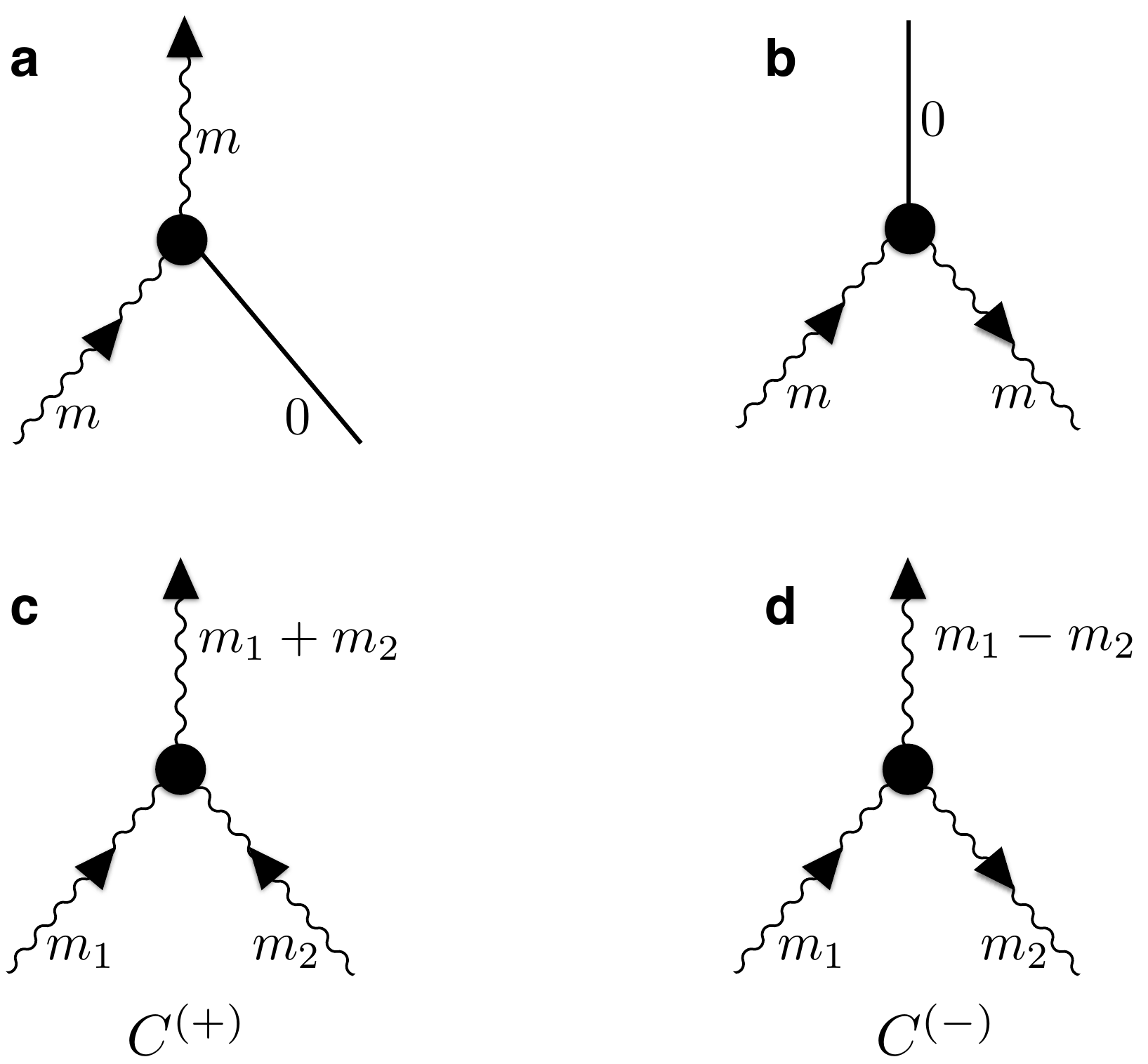}
\caption{Triad interactions organized by zonal wavenumbers.  (a) The wave -- mean-flow interaction in which an eddy of zonal wavenumber $m$ interacts with the zonal mean flow.  (b) Reynolds stress on the zonal mean-flow induced by two waves of equal but opposite zonal wavenumber.   Processes (a) and (b) are the only nonlinearities included in the quasi-linear (QL) and second-order cumulant (CE2) approximations. (c) General triad interaction involving two waves with positive zonal wavenumbers.  This scattering process has amplitude $C^{(+)}_{\ell; \ell_1 m_1; \ell_2 m_2}$.  (d) General triad interaction involving one wave with positive zonal wavenumber and another wave with negative wavenumber and amplitude $C^{(-)}_{\ell; \ell_1 m_1; \ell_2 m_2}$.  Scatterings (c) and (d) are included in DNS and in CE2.5 and CE3.}
\label{triads}
\end{figure}

Spherical harmonics are the most convenient basis for models on the sphere that possess zonal symmetry.
In this representation of the dynamical variables the complex coefficients $\{ q_{\ell m} (t)\}$ are defined by the spectral expansion
\begin{eqnarray}
q(\theta, \phi, t) \approx \sum_{\ell = 0}^L \sum_{m = -\min\{\ell, M\}}^{\min\{\ell, M\}} q_{\ell m}(t)~ Y_{\ell m}(\theta, \phi),
\end{eqnarray}
with spherical wavenumber cutoff $L$ and zonal wavenumber cutoff $M$. Here $\theta$ and $\phi$ are co-latitude and longitude respectively, and the spherical harmonics $Y_{\ell m}$ are defined such that $Y_{\ell, -m} = Y_{\ell m}^{*}$.
Real-valuedness of the fields then manifests itself as
\begin{eqnarray}
q_{\ell, -m} = q_{\ell m}^{*},
\end{eqnarray}
so it suffices to focus only on the evolution of modes with zonal wavenumber $m \geq 0$.
 Their time evolution is governed by the spectral representation of the EOM that has the form
\begin{eqnarray}
\dot{\zeta}_{\ell m} &=& A_{\ell}~ \delta_{m,0} + \sum_{\ell_1} B_{\ell; \ell_1 m}~ \zeta_{\ell_1 m} + \eta_{\ell m}(t)
\nonumber \\
&+& \sum_{\ell_1, \ell_2, m_1, m_2}^{m = m_1+m_2} C^{(+)}_{\ell; \ell_1 m_1; \ell_2 m_2}~  \zeta_{\ell_1 m_1} \zeta_{\ell_2 m_2} 
\nonumber \\
&+&  \sum_{\ell_1, \ell_2, m_1, m_2}^{m = m_1-m_2} C^{(-)}_{\ell; \ell_1 m_1; \ell_2 m_2}~  \zeta_{\ell_1 m_1} \zeta^*_{\ell_2 m_2}\ .
\label{FluidspectralEOM}
\end{eqnarray}
Coefficients $A_\ell$ and $B_{\ell; \ell_1 m}$ are the matrix elements of the constant and linear operators (including the Coriolis term) that appear in the EOM.  ($A_\ell = 0$ for the jet problem that we study here.)  The quadratic nonlinearities have their origin in the Jacobian with coefficients $C^{(+)}$ representing amplitudes for the scattering of two waves each with zonal wavenumber $m \geq 0$; $C^{(-)}$ are for waves with $m > 0$ and $m < 0$ to scatter (Figure \ref{triads}). The added complexity of separating the quadratic nonlinear term into two parts is the price to pay for the convenience of only focusing on modes with $m \geq 0$.
The coefficients are obtained (and stored) from the matrix elements of the Jacobian:  $J^{(\pm)}_{\ell; \ell_1 m_1; \ell_2 m_2} = I^{(\pm)}_{\ell; \ell_1 m_1; \ell_2 m_2} \mp 
I^{(\pm)}_{\ell; \ell_2 m_2; \ell_1 m_1}$ where the integrals $I^{(\pm)}$ are over products of associated Legendre functions:
\begin{eqnarray}
I^{(\pm)}_{\ell; \ell_1 m_1; \ell_2 m_2} &\equiv& 2 \pi i~ m_1 \int_0^\pi P_{\ell}^{m_1 \pm m_2}(\cos \theta)~ P_{\ell_1}^{m_1}(\cos \theta)
\nonumber \\
&\times& \frac{\partial}{\partial \theta} P_{\ell_2}^{m_2}(\cos \theta)~ d\theta\ .
\end{eqnarray}
Integrals $I^{(\pm)}$ may be evaluated in a numerically exact manner by Gaussian quadrature, or by relating them to 6j-symbols \citep{silberman1954,thiebaux1971}.   
The total zonal wavenumber of each term on the right-hand side of Equation~(\ref{FluidspectralEOM}) equals $m$, which is derived from integration $\int_0^{2\pi} d\phi$ of a product of two or three spherical harmonics, reflecting the zonal symmetry of dynamics on the sphere. These are the equations to be solved for a purely spectral Direct Numerical Simulation (DNS; see Section~\ref{DNS}).

This spectral approach is also utilized in Direct Statistical Simulation. On symmetry grounds, the first zonally-averaged cumulant must be independent of longitude $\phi$ and can be written $c(\theta)$. Therefore in the spherical harmonic basis only the $m = 0$ mode of the first cumulant, $c_\ell = \zeta_{\ell, m=0}$, is non-zero. Similar symmetry arguments yield the result that the second cumulant depends on the latitudes of the two field points, but only on the difference between their longitudes, $c(\theta_1, \theta_2, \phi_1 - \phi_2)$. It can therefore be written in the spherical harmonic basis as $c_{\ell_1 \ell_2 m} =  \zeta_{\ell_1 m}~ \zeta_{\ell_2 -m}  - c_{\ell_1} c_{\ell_2} \delta_{m 0} = \zeta_{\ell_1 m}~ \zeta_{\ell_2 m}^*  - c_{\ell_1} c_{\ell_2} \delta_{m 0}$. Note that $c_{\ell_1 \ell_2 ,m=0} = 0$. Moreover,
because the scalar fields are real-valued in coordinate space, and $c(\theta_1, \theta_2, \Delta \phi) = c(\theta_2, \theta_1, -\Delta \phi)$ we have $c_{\ell_1 \ell_2 m} = c_{\ell_2 \ell_1 m}^*$.  
At order CE2, the equations of motion for the first two orders of cumulants in this basis read:
\begin{eqnarray}
\dot{c}_\ell &=& A_\ell + \tilde{B}_{\ell; \ell_1 0}~ c_{\ell_1} + C^{(-)}_{\ell; \ell_1 m; \ell_2 m}~  c_{\ell_1 \ell_2 m}
\label{ceom1}
\end{eqnarray}
and
\begin{eqnarray}
\dot{c}_{\ell_1 \ell_2 m} &=& 2 \Gamma_{\ell_1 \ell_2 m} + \left\{ 2 \tilde{B}_{\ell_1; \ell m}~ c_{\ell \ell_2 m} \right\}
\label{ceom2}
\end{eqnarray}
where 
\begin{eqnarray}
\tilde{B}_{\ell_1; \ell m} \equiv B_{\ell_1; \ell m} + C^{(+)}_{\ell_1; \ell^\prime 0; \ell m}~ c_{\ell^\prime}
\label{Bprime}
\end{eqnarray}
are the linear terms in the EOM, renormalized by the mean-flow.  
The convention of summation over repeated indices is adopted here, and in the spectral basis $\left\{ c_{\ell_1 \ell_2 m} \right\} \equiv \frac{1}{2} (c_{\ell_1 \ell_2 m} + c^*_{\ell_2 \ell_1 m})$.  
Spectral power of relative vorticity in a mode with spherical wavenumber $\ell$ and zonal wavenumber $m$ is given by $|\zeta_{\ell m}|^2 = c_{\ell \ell m} + c_{\ell}^2 \delta_{m 0}$.  Likewise integrals of motion such as the total angular momentum, kinetic energy, and enstrophy can be easily obtained from the first and second cumulants.  

Due to zonal symmetry the third cumulant is a function of only 5, not 6, wavenumbers, and can be written as  $c_{\ell; \ell_1 m_1; \ell_2 m_2} \equiv \zeta_{\ell, m_2 - m_1}~ \zeta_{\ell_1 m_1}~ \zeta^*_{\ell_2 m_2} $ under the assumption that zonal wave vectors $m_1 \neq 0$, $m_2 \neq 0$, and $m_1 \neq m_2$. 
The third cumulant vanishes by zonal symmetry if $m_1 = 0$, or $m_2 = 0$, or $m_1 = m_2$.
At order CE3, Equation~(\ref{ceom2}) is supplemented with:
\begin{eqnarray}
\dot{c}_{\ell_1 \ell_2 m} &=& \cdots + \bigg{\{} C^{(+)}_{\ell_1; \ell m-m_1; \ell_1^\prime m_1} c_{\ell; \ell^\prime_1 m_1; \ell_2 m}
\nonumber \\
&+& C^{(-) *}_{\ell_2; \ell m+m_1; \ell_1^\prime m_1} c_{\ell_1; \ell^\prime_1 m_1; \ell m+m_1} \bigg{\}}\ .
\label{3rdTo2}
\end{eqnarray}
The EOM for the third cumulant is given in the spectral basis by:
\begin{eqnarray}
\dot{c}_{\ell; \ell_1 m_1; \ell_2 m_2} &=& \bigg{\{} 2 \tilde{B}_{\ell_1; \ell_1^\prime m_1}~ c_{\ell; \ell_1^\prime m_1; \ell_2 m_2}
\nonumber \\
&+& \tilde{B}^*_{\ell_2; \ell^\prime_2 m_2}~ c_{\ell; \ell_1 m_1; \ell^\prime_2 m_2}
\nonumber \\
&+& 2 C^{(+) *}_{\ell_2; \ell_1^\prime m_1; \ell_2^\prime m_2-m_1}~ c_{\ell_1, \ell^\prime_1, m_1}~ c_{\ell, \ell^\prime_2, m_2-m_1} 
\nonumber \\
&+& 2 C^{(-)}_{\ell; \ell_1^\prime m_1; \ell_2^\prime m_2}~ c_{\ell^\prime_1, \ell_1, m_1}~ c_{\ell_2, \ell^\prime_2, m_2}\bigg{\}}
\nonumber \\
&-& \frac{1}{\tau}~ c_{\ell; \ell_1 m_1; \ell_2 m_2}
\label{ceom3}
\end{eqnarray}
with $\left\{c_{\ell; \ell_1 m_1; \ell_2 m_2} \right\} \equiv \frac{1}{2} (c_{\ell; \ell_1 m_1; \ell_2 m_2} + c_{\ell_1; \ell m_2 - m_1; \ell_2 m_2})$.  For the special case $m_2 = 2 m_1$ the third term on the RHS of Equation \ref{ceom3} has coefficient $C^{(+) *}$ (rather than $2 C^{(+) *}$); this has been suppressed for the sake of clarity.  

The projection operator Equation~(\ref{projection}) of CE3$^*$ is implemented in spectral space by diagonalizing, for each value of zonal wavenumber $m$, the $L \times L$ Hermitian matrix $c_{\ell_1 \ell_2 m}$, removing all eigenvectors with negative eigenvalues, and then rebuilding the second cumulant from the remaining eigenvectors and values.      

A diagnostic equation for the third cumulant in the simplest CE2.5 approximation can now be obtained from Equation~(\ref{ceom3}) by setting the left-hand side to zero, and dropping the two terms involving $\tilde{B}$ to avoid having to solve a non-trivial linear 
equation.  The third cumulant in this CE2.5 approximation is then determined by products of two second cumulants:
\begin{eqnarray}
c_{\ell; \ell_1 m_1; \ell_2 m_2} &=& 2 \tau \bigg{\{} 
C^{(+) *}_{\ell_2; \ell_1^\prime m_1; \ell_2^\prime m_2-m_1}~ c_{\ell_1, \ell^\prime_1, m_1}~ c_{\ell, \ell^\prime_2, m_2-m_1}  
\nonumber \\
&+& C^{(-)}_{\ell; \ell_1^\prime m_1; \ell_2^\prime m_2}~ c_{\ell^\prime_1, \ell_1, m_1}~ c_{\ell_2, \ell^\prime_2, m_2}\bigg{\}}\ .
\label{3rdcum}
\end{eqnarray}
Substitution of Equation~(\ref{3rdcum}) into Equation~(\ref{3rdTo2}) closes the EOM at the CE2.5 level of approximation.  

\subsection{Symmetry of Reflection About Equator}

Many model geophysical and astrophysical systems are governed by dynamical equations that have definite north-south symmetry of reflection about the equator.  Assuming ergodic behavior, we expect statistics accumulated in time to reflect this symmetry.  Zonal means, on the other hand, can either be steady at long times or oscillate.  The fixed point that describes the former case should also have definite symmetry, but oscillating zonal means will generally only have north-south symmetry if they are time-averaged.  We will see examples of both types of behavior below.  

Often CE2 exhibits oscillations in the zonal means not exhibited by the full dynamics that we may wish to partly suppress by imposing equatorial reflection symmetry directly upon the statistics.  This can be done by requiring the first and second cumulants to possess definite symmetry.  For the first cumulant we have in real space $c(\theta) = \pm c(\pi - \theta)$ or in the basis of spherical harmonics $c_\ell = 0$ for $\ell$ odd (even).  The later relation follows from the symmetry of spherical harmonics: $Y^\ell_m(\pi - \theta, \phi) = (-1)^{\ell+m}~ Y^\ell_m(\theta, \phi)$.  The second cumulant likewise obeys the rule: $c_{\ell_1 \ell_2 m} = 0$ if $\ell_1 + \ell_2$ is odd.  

\subsection{Initial Conditions}

The initial state for DNS and QL DNS is usually taken to be at rest, $\zeta = 0$, but other initial states are investigated as discussed below.  For DSS, the second cumulant is initialized to have only local-in-space correlations:  $c_{\ell_1, \ell_2, m} = c~ \delta_{\ell_1, \ell_2}$ where $c$ is a small positive constant.  This imparts minimal bias to the subsequent evolution as any flow must have positive autocorrelations at a single space-time point.  It is also possible to initialize $c_{\ell_1, \ell_2, m} = 0$ as by Equation \ref{ceom2} the stochastic forcing will generate correlations.  
The first cumulant is initialized in two different ways.  To ensure the formation of jet centered on the equator we may set all components to zero except for ${\ell = 3}$.  Then, depending on the sign of $c_{\ell = 3}$ either a prograde or retrograde jet is encouraged to form.  Alternatively, the $c_\ell$ may be initialized with random amplitudes to study the possible existence of multiple fixed points.  For CE3$^*$ the third cumulant is always initialized to be zero.  

\subsection{Numerical Implementation of DNS}
\label{DNS}

Pure spectral DNS with truncation $0 \leq \ell \leq L$ and $|m| \leq \min\{\ell, M\}$ is performed.  We choose spectral cutoffs $L = 30$ and $M =20$ and demonstrate below by comparison to a high-resolution simulation that these cutoffs suffice.  
We work on the unit sphere and in units of time such that the Coriolis parameter $f = 2 \Omega \cos(\theta)$ with $\Omega = 2 \pi$.  
To remove enstrophy cascading to small scales, hyperviscosity $\nu_3 (\nabla^2+2) \nabla^4 \zeta$ is included in the linear operator of Equation~(\ref{EOMJ}).   
The coefficient $\nu_3$ is chosen such that the most rapidly dissipating mode decays at a rate of $1$.
The pure spectral EOMs of Equation~(\ref{FluidspectralEOM}) are integrated forward in time using a fourth-order-accurate Runge-Kutta algorithm with an adaptive time step $\Delta t$.  Each time step requires ${\cal O}(L^3 M^2)$ floating point computations that at high resolutions would be prohibitively expensive compared to a pseudo-spectral algorithm but is feasible here for the moderate resolutions that we study.  The calculation of a time step is made faster by skipping over triads that vanish due to symmetry.  At each time step, the stochastic forcing is updated to a new value by the following scheme \citep{lilly1969}:
\begin{eqnarray}
\eta_{n+1}  =  R~ \eta_{n} + \sqrt{1- R^2}~ \hat{\eta}_{n+1}, 
\end{eqnarray}
where $\eta_n$ is the stochastic forcing at time step $n$, the memory coefficient $R = (1- \Delta t / \tau_r)/(1+\Delta t / \tau_r)$ with stochastic renewal time $\tau_r$, and each of the real and imaginary parts of the complex number $\hat{\eta}_{n+1}$ is randomly drawn from a Gaussian distribution with zero mean.   For eddy-turnover timescales that are much larger than the stochastic renewal timescale the forcing obeys Gaussian statistics with zero mean and approximate space-time correlations
\begin{eqnarray}
\langle \eta_{\ell m}(t)~ \eta^*_{\ell^\prime m^\prime}(t^\prime) \rangle  = 2 \Gamma_{\ell \ell^\prime m}~ \delta_{m m^\prime}~ \delta(t - t^\prime)\ . 
\end{eqnarray} 

Quasi-linear (QL) DNS is performed by calculating the dynamics in a reduced model where only the eddy-scattering triad interactions corresponding to Figure~\ref{triads} (a) and (b) are included \citep{herring1963,ogormanschneider2007}, cutting the cost of calculating a time step to ${\cal O}(L^3 M)$.  \citet{srinivasanyoung2012} studied the jet formation problem on the $\beta$-plane and termed the removed interactions as the EENL --- the eddy-eddy nonlinearity.  Time-average statistics obtained from QL DNS should be the same as those obtained from CE2 as the CE2 closure is exact in the case of QL dynamics.   This is the reason why CE2 is a realizable closure.  We note that the third cumulant is generally non-zero in QL DNS; however it decouples from the first and second cumulants and does not contribute to their tendencies.  

To verify that the full spectral simulation has sufficient resolution, finer-scale DNS of the fluid is also performed in real space on a spherical geodesic grid \citep{heikesrandall1995a,heikesrandall1995b,qimarston2014} of D = 163,842 cells; the lattice operators conserve energy and enstrophy. The vorticity evolves forward in time by a second-order accurate leapfrog algorithm, with a Robert-Asselin-Williams filter of $0.001$ and $\alpha = 0.53$ \citep{williams2009}.  The time step is fixed at $\Delta t = 0.003$.  

A program that implements spectral DNS, real-space DNS, and DSS and includes all the graphical tools needed to visualize statistics, is freely available\footnote{The application ``GCM'' is available for OS X 10.9 and higher on the Apple Mac App Store at URL http://appstore.com/mac/gcm}.  The Objective-C++ and Swift programming languages are employed. C blocks and Grand Central Dispatch enable the efficient use of multiple CPU cores.  

\subsection{Numerical Implementation of DSS}
\label{numericalDSS}

To study properly the accuracy and predictive power of DSS, it is necessary that it be applied to precisely the same model that is simulated by DNS.   Object-oriented programming makes this straightforward.  The cumulant expansions are implemented as subclasses of the same spectral class that implements DNS, re-using the same methods.  Any differences between DNS and DSS can therefore be ascribed to the different approximate closures.

Because the 2nd and 3rd cumulants have higher dimension (3 and 5 respectively) than the dynamical fields (dimension 2) the equations of motion for CE2.5 and CE3$^*$ are computationally demanding.  Factorization of products of 3 matrices, however, into 2 separate products of 2 matrices reduces the computational burden by a factor of $L$.  The result is that a time step of CE2 requires ${\cal O}(L^3 M)$ operations, while CE2.5 and CE3$^*$ require ${\cal O}(L^4 M^2)$.  The projection operator, Equation~(\ref{projection}), only requires ${\cal O}(L^3 M)$ operations.  It turns out that pseudo-spectral algorithms offer no advantage for DSS on the sphere, as it requires the same order of operations and in fact may be slower due to large prefactors \citep{tobiasdagonetal2011}.  

For the results presented here, explicit time integration of the EOMs for the cumulants is again done with the 4th-order accurate Runge-Kutta algorithm with adaptive time step.  Because the EOMs for the cumulants tend to be stiff, other algorithms can be faster.  We have implemented two other approaches.  Implicit time integration by backwards differentiation (BDF1 and BDF2) using the method of Krylov subspaces \citep{saad2003} can be much faster than explicit integration when DSS reaches a fixed point.   We have also implemented a fixed point method that directly solves the time-independent Lyapunov equation for the second cumulant, Equation~(\ref{ceom2}) with LHS = 0.  At present, however, the method only works for stochastically-driven systems that have no neutral modes (see Section~\ref{results} below).  We will report details of these methods elsewhere.

\section{Illustrative jet problem and the zonostrophy parameter}
\label{jetProblem}

Of course there are many examples of direct numerical simulation of barotropic jet problems both on the $\beta$-plane and a spherical surface. Usually each of these numerical simulations is designed to highlight one particular aspect of the problem, for example the importance of the choice of driving terms \citep{scottdritschel2012} or the importance of the dissipation term in determining whether equatorial super-rotation or sub-rotation is preferred \citep{warneforddellar2013,scottpolvani2008}. We shall not summarize the important results that have been achieved using this method here, since they are described in other Chapters. What interests us here is determining when the statistics obtained by DSS give an accurate description of those achieved by averaging the results from DNS. Clearly it is not always expected that truncations of DSS give good approximations to the true statistics --- for example quasilinear approximations may not always be appropriate. 

It turns out that, for the jet formation problem, an important parameter determining the efficacy of DSS at various truncations is the zonostrophy parameter \citep{galperinsukorianskyetal2006} --- see below for a precise definition.  DNS of jet formation on the $\beta$-plane reveals the importance of this parameter for the dynamics and hence the statistics of the system.  As described in detail elsewhere in this Volume, the physical
interactions underpinning the formation and evolution of jets on the
$\beta$-plane have been studied in great detail using both theoretical
arguments and Direct Numerical Simulation. We therefore do not give a
complete review here, but mention only the limited dynamics and interactions
relevant to our investigation of the efficacy of DSS; for our purposes
it is natural to identify the typical lengthscales and timescales that are
important for jet formation and discuss the fidelity of DSS as the
ratios between these lengthscales and timescales are changed in the problem.

An important lengthscale can be identified immediately from
Equation~(\ref{EOMJ}) by calculating at what scales the linear and
nonlinear parts of the Jacobian operator (i.e. the inertial term and
the $\beta$-effect term) are comparable.  This scale can also be associated with that at which zonal
flows become important in mediating the dynamics of  propagating
nonlinear Rossby waves (see for instance \citet{rhines1975,rhines1979,vallis2006} and also the
Chapter by Bouchet, Nardini and Tangarife in this Volume).  This ``Rhines scale'' is given by
$L_R=(2U/\beta)^{\frac{1}{2}}$, where $U$ is the rms velocity of the
flow. For unforced, non-dissipative flows this is the only lengthscale
that may play a role. However, when energy is input into the system
(via a driving term) and dissipated either by friction or viscous
effects, other lengthscales may become important. For
example it is conceivable that both the scale of forcing and that of
dissipation might play a role; the former being more likely than the
latter to influence the large-scale dynamics. However, if the forcing
scale is far removed from the large scales, it is more likely that the
scale measuring the relative strengths of the forcing to the
background potential vorticity gradient will be important. This
lengthscale, originally introduced in \citet{maltrudvallis1991} and
denoted by $L_\varepsilon$ is now  thought to be important in the
dynamics of zonation. For the simple $\beta$-plane model  $L_\varepsilon=
2 (\varepsilon/\beta^3)^{\frac{1}{5}}$ where $\varepsilon$ is the energy input
rate of the stochastic forcing $\eta$. \citet{vallismaltrud1993} discuss how the
presence of a mean gradient of potential vorticity can selectively
(i.e. anisotropically) inhibit the cascade of energy from the forcing
scale to large scales. They derived the form of a  region of wavenumber space in
which the energy arriving from small scales is significantly
suppressed, owing to the selective nature of the triad interactions of
Rossby waves. This ``dumb-bell'' region can be used to explain the
formation of zonal flows, since energy transfer is not forbidden into
a thin strip perpendicular to the zonal direction.

Ignoring for the moment the possible importance of the driving and
dissipation scales, we expect the ratio of the Rhines scale ($L_R$)
to  $L_\varepsilon$ to play an important role in determining the
dynamics of the jets. This ratio is termed the
zonostrophy parameter and is given by $R_\beta
\equiv L_R / L_\varepsilon = U^{1/2} \beta^{1/10}/(\sqrt{2}~ \varepsilon^{1/5})$ \citep{galperinsukorianskyetal2010}.
If $R_\beta$ is small the forcing dominates the natural variability of
the system; the scale at which the system forgets that it is forced is
larger than the Rhines scale; here the jets are weak, meander significantly
and no staircase is formed \citep{scottdritschel2012}. However if the zonostrophy
index is large  then stable jets are found as the forcing is not strong enough to knock the natural
dynamics out of equilibrium. The zonostrophy parameter is therefore a measure of how far the system is driven out of
equilibrium, a fact that becomes clearer when $R_\beta$ is written in terms
of the ratio of an advective time on the Rhines scale to a dissipative timescale as noted by
\citet{tobiasmarston2013}. The quasi-equilibrium limit is therefore given by $R_\beta
\rightarrow \infty$. We note here that it has been demonstrated that  in certain
circumstances the forcing lengthscale remains important; then
the dynamics is controlled by two non-dimensional parameters
separately \citep{srinivasanyoung2012,bouchetnardinietal2013}. Even in this regime, however
we believe that $R_\beta$ does give {\it a} measure of
the degree of lack of equilibrium. We shall therefore
calibrate how well DSS performs as a function of this ratio.

We choose parameters for the jet as in \citep{tobiasdagonetal2011} 
such that it is moderately far away from equilibrium and within a realistic range for real planetary atmospheres.  
The fluid motion is driven by stochastic forcing $\eta$ and damped by friction $\kappa$.  The friction parameter $\kappa = 0.02$ and thus the friction relaxation time scale is $50$.   Only modes with $8 \leq \ell \leq 12$ and $8 \leq |m| \leq \ell$ are stochastically forced (with $|m| \leq \ell$).  This has the effect of confining the stochastic forcing to lower latitudes, enabling several illuminating numerical experiments.  For instance fluid motion that is excited near the poles is a diagnostic of eddy-eddy scattering that sends momentum into high latitudes.   Note that the forcing has no $\ell = 0$ or $\ell = 1$ modes, so the zero circulation constraint is preserved and no net angular momentum is injected.  We set the stochastic driving force to be $\Gamma_{\ell \ell^\prime m} = 0.1 \delta_{\ell \ell^\prime}$ for $8 \leq \ell \leq 12$ and $8 \leq |m| \leq \ell$
and set the stochastic renewal time to be $\tau_r = 0.1$.  The time step, though adjustable, is constrained to be small compared with the stochastic renewal time:  $\Delta t < 0.1 \tau_r$.  For quasi-linear DNS it is necessary to apply the tighter constraint of $\Delta t < 0.03 \tau_r$ to reproduce accurately the jet orientation found in the other approaches.  The kinetic energy density for all simulations is approximately $e = E/4 \pi \approx 0.042$ giving $U = \sqrt{2 e} \approx 0.29$ and an eddy turnover timescale of order $T = 1/U \approx 3.5$; thus there is an adequate separation of time scales with $\Delta t \ll \tau_r \ll T$.

The Rhines scale may be estimated by using the value of $\beta = 4 \pi \sin(\theta)$ on the equator ($\theta = \pi/2$) where the prograde jet is centered; this combined with U gives $L_R \approx 0.215$.   Further estimating $\varepsilon = \kappa U^2$ yields zonostrophy parameter $R_\beta \approx 1.76$.  (If $\beta$ at a latitude of $45^\circ$ is used instead, $R_\beta \approx 1.7$ -- only slightly less because of the smallness of the exponent $1/5$.)  The zonostrophy parameter is therefore between those of the $R_\beta \approx 1.98$ and $R_\beta \approx 1.24$ jets studied on the $\beta$-plane by \citet{tobiasmarston2013}.  It exhibits eddies large enough to provide an interesting test of the hierarchy of CE closures.  We emphasize, however, that the setup of the problem on the sphere differs from that on the $\beta$-plane in important respects.  The sphere lacks translational symmetry in the meridional direction, and as discussed above we choose the stochastic forcing to be strongest around the equator rather than homogeneous in space as in \citet{tobiasmarston2013}.  As explained below in Section \ref{results} this setup of the numerical experiment permits us to probe weaknesses in DSS that were not apparent on the $\beta$-plane.  It also drives home the point that a single diagnostic such as the zonostrophy parameter does not fully specify the behavior of a jet.

\subsection{DNS of Jet}
\label{JetDNS}

Our philosophy is that much is learned about the technique of DSS when it {\em fails} to reproduce the results obtained from the long-term averaging of DNS. We therefore believe it is crucial to test statistical approaches against full DNS {\it for precisely the same model and at precisely the same resolution and same parameters}. This was first attempted for the case of the stochastically driven jet problem (with and without magnetic field) by \citet{tobiasdagonetal2011}.

\begin{figure*}
\figurebox{25pc}{}{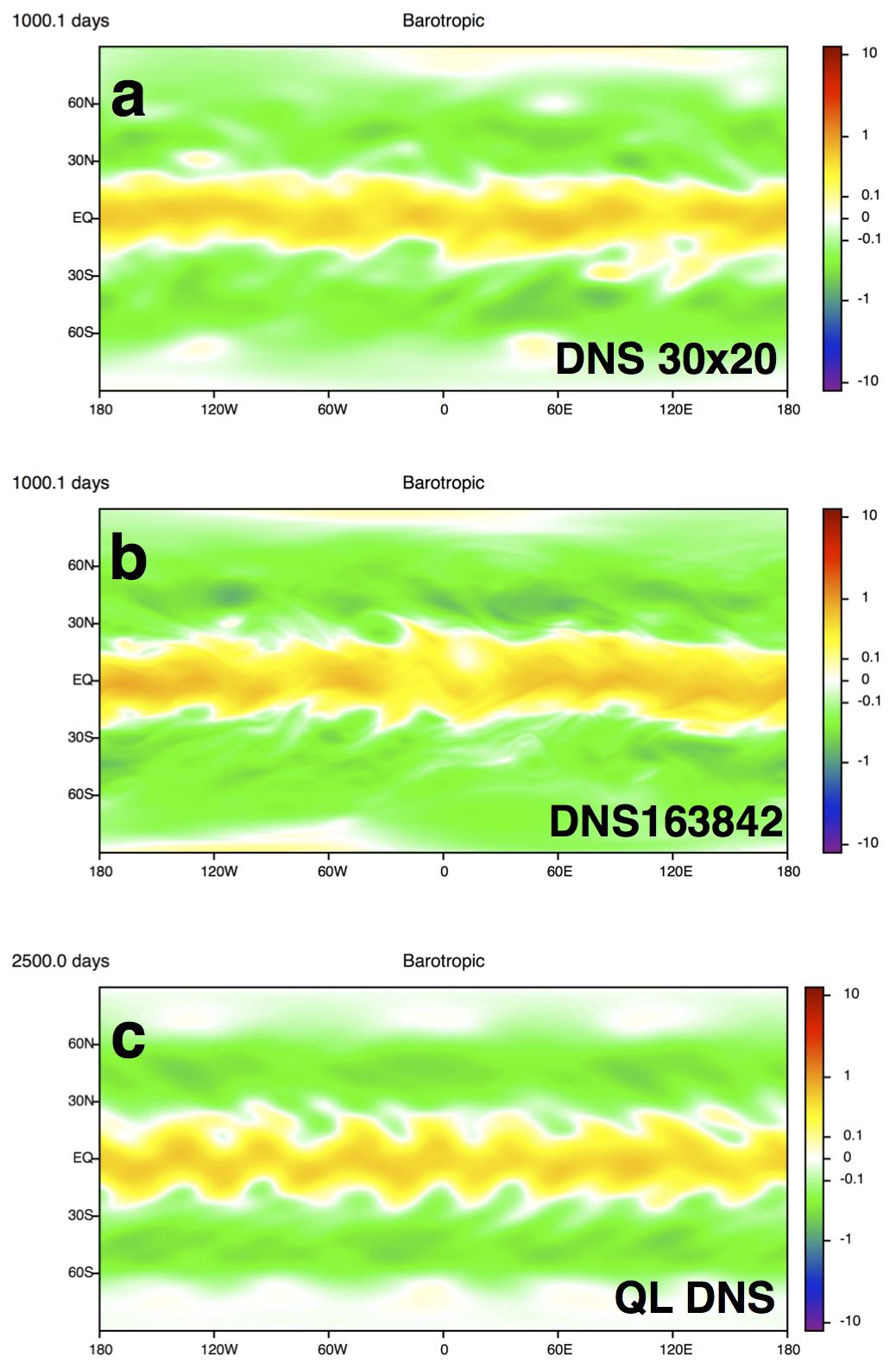}
\caption{Instantaneous snapshot of the zonal velocity as obtained from DNS after spin-up.   (a) Spectral simulation with truncation $0 \leq \ell \leq L$ for $L = 30$ and $|m| \leq \min\{\ell, M\}$ for $M = 20$.  (b) Spherical geodesic grid with 163,842 cells.  (c) Quasi-linear DNS for the same spectral truncation as (a).}
\label{dns}
\end{figure*}

DNS in both spectral space and real space show the spontaneous formation of three coherent zonal jets.  Figure \ref{dns} shows snapshots of the instantaneous zonal velocity field at time $t = 1000$.  By that time the system has reached a statistically steady state where energy injection from the random forcing balances energy dissipation.  There is an eastward (prograde) jet centered on the equator, and two westward (retrograde) mid-latitude jets on either side.  The total angular momentum is close to zero.  The spectral simulation and real-space simulation agree well with each other.   No particular significance should be attached to the super-rotation; different choices of the stochastic forcing can lead to either prograde or retrograde equatorial jets.  

As is well-known, the dynamics of this system is such that energy that is injected into the system at moderate scales is transported by eddies to larger scales. The interaction with the gradient in planetary vorticity leads to the large-scale dynamics being anisotropic and the formation of zonal jet structures.  That the quasi-linear spectral simulation, with no mechanisms for cascades, can exhibit qualitatively the same coherent jets directly demonstrates that a scale-by-scale cascade mechanism is not a requirement for jets to form.  Nevertheless it is evident from Figure \ref{dns} (c) that the quasilinear jet also has more coherent waves than fully nonlinear DNS.  As shown below this is also evident in the CE2 simulations. 

\subsection{DSS Compared with DNS}
\label{results}

In \citet{tobiasmarston2013} we demonstrated that DSS truncated at CE2 gives an accurate description of the statistics of $\beta$-plane turbulence when the system is close to equilibrium as measured by the zonostrophy parameter. However as the system is driven further from equilibrium the neglect of the eddy-eddy scattering has implications for the accuracy of this quasilinear DSS. As the zonostrophy parameter is decreased the method first of all fails to predict accurately the form of the second cumulant. The strict quasilinear truncation appears to fail to describe the appearance of ``satellite modes'' --- these are modes of zonal wavenumber one that play a part in mediating the dynamics of the system. This occurs even when the truncation gives a good description of the first cumulant (i.e.\ the mean flow). Further decrease in the zonostrophy parameter means that the system is further from equilibrium and the jets are more intermittent and meander more. In this regime CE2 not only fails to reproduce the form of the second cumulant, but also the number and strength of the zonal jets. These results serve as motivation for the calculations presented here. We investigate whether the inclusion of eddy-eddy scattering in the cumulant expansion can lead to a more accurate description of the low-order cumulants. In this case we investigate the formation of jets in QL DNS, CE2, CE3$^*$ and CE2.5 on a spherical barotropic surface.

\begin{figure*}
\figurebox{45pc}{}{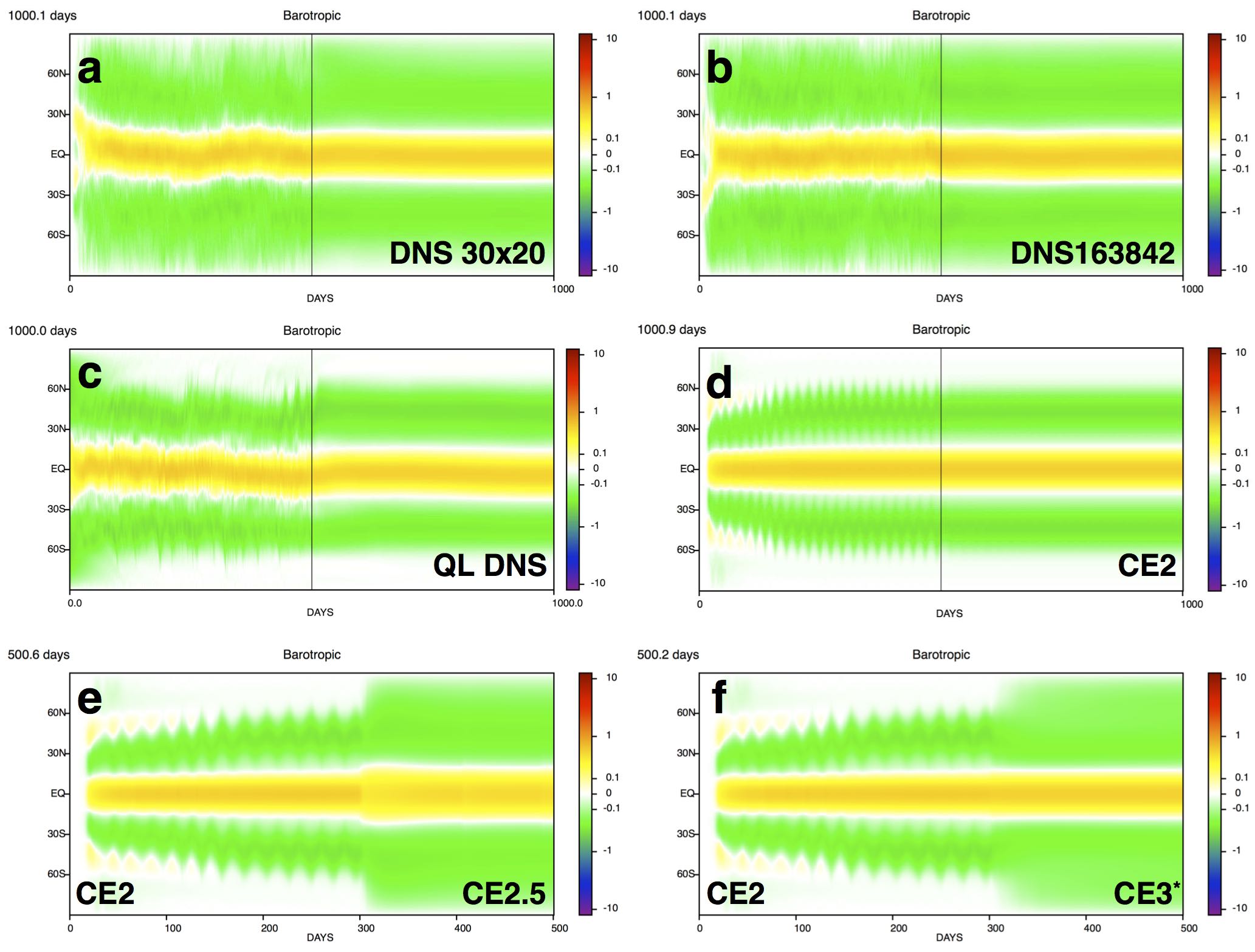}
\caption{Hovm\"oller timelines of the zonal mean zonal velocity.  Time averaging commences at the times indicated by the vertical black line in (a), (b), (c) and (d).  (a) Spectral simulation with truncation $0 \leq \ell \leq L$ for $L = 30$ and $|m| \leq \min\{\ell, M\}$ for $M = 20$.  (b) DNS on spherical geodesic grid with 163,842 cells.  (c) Quasi-linear DNS for the same spectral truncation as (a).  (d) CE2.  (e) CE2.5 with $\tau = 2$.  For times $t < 300$ only CE2; the CE2.5 correction is turned on at $t = 300$.  (f)  CE3$^*$ with $\tau = \infty$ (CE2 for $t < 300$).   The QL DNS and CE2 simulations are initialized such that they produce a prograde equatorial jet.  (See Figure \ref{reversal} for retrograde jets.)  All the cumulant expansions have north-south reflection symmetry about the equator imposed and have precisely the same spectral truncation and parameters as (a) and (c).}
\label{timelines}
\end{figure*}

Figure \ref{timelines} presents Hovm\"oller timelines of the zonal mean zonal velocity.   Spectral and real-space DNS simulations agree well.  The retrograde flow persists to high latitudes, unlike the QL DNS or CE2 simulations, which lack a mechanism to scatter eddies and their associated angular momentum from low latitudes where the stochastic forcing is strongest to high latitudes.  The high-latitude retrograde flows are, however, captured by CE2.5 and CE3$^*$ as some physics of the eddy-eddy interaction is captured at these levels of approximation.  Unphysical slow oscillations in the zonal mean exhibited by CE2 are damped out once the eddy-eddy interactions are turned on at time $t = 300$.  A running time average commences at the times marked by the vertical black line (CE2.5 and CE3$^*$ reach a stable fixed point and time-averaging is not needed).   We note that time-averaging does not commute with zonal averaging for two-point and higher order statistics because zonal means can fluctuate with time.  For example, consider the 2nd cumulant.  If zonal averaging is performed first then by construction there is no $m = 0$ component, and that continues to hold upon time averaging.  If time averaging $\langle \rangle$ is performed first, however,  $\langle (\zeta_{\ell, m=0})^2 \rangle > (\langle \zeta_{\ell, m=0} \rangle)^2$, and the second cumulant generally acquires a positive $m=0$ component that persists upon zonal averaging.  Here we only time average after first performing the zonal average.

\begin{figure*}
\figurebox{45pc}{}{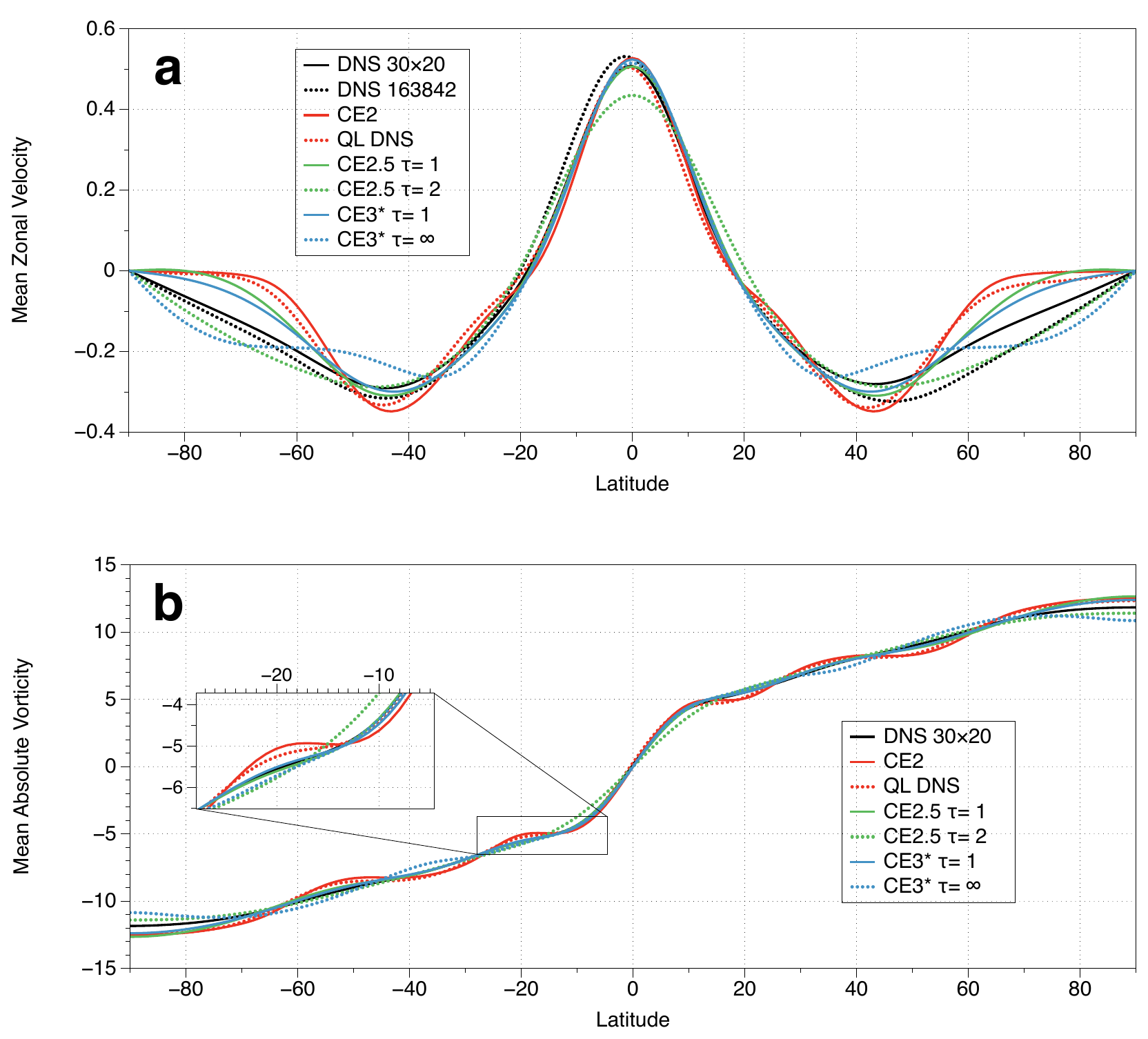}
\caption{Comparison of zonal means as a function of latitude as calculated by DNS and DSS.  (a) Zonal velocity.  It can be seen that QL DNS and CE2 do not scatter enough angular momentum into high latitudes where the stochastic forcing is weak.  CE2.5 with $\tau = 2$ scatters excessively and suppresses the jet around the equator.  CE3$^*$ with no eddy damping scatters too much momentum to high latitudes.  (b) Absolute vorticity.  CE2 and QL DNS exaggerate the steps in the vorticity.}
\label{zonalMeans}
\end{figure*}

A quantitative comparison of the zonal means is presented in Figure \ref{zonalMeans}.  The spectral and real-space DNS simulations agree closely, demonstrating that the spectral truncation is sufficient.  The small deviation from north-south symmetry of reflection about the equator seen in the real-space DNS simulation is due to slight breaking of that symmetry by the spherical geodesic grid.   Likewise CE2 and QL DNS are very close as expected, and qualitatively capture the three-jet coherent structure but with insufficient flow at high latitudes.   Higher-order CE2.5 and CE3$^*$ closures agree better quantitatively with DNS.  While the results vary depending on the value of $\tau$ ($= 1$, $2$, or $\infty$) the relative insensitivity to the choice of $\tau$ demonstrates that DSS has predictive power even when an eddy-damping parameter is used.  CE3$^*$ at $\tau = \infty$ exaggerates the influence of eddies compared to DNS.  The plot of the zonal mean absolute vorticity reveals a rounded staircase instead of the sharp plateaus found for weakly driven jets much closer to equilibrium \citep{scottdritschel2012}.  CE2 exaggerates the sharpness of the steps.  
 
\begin{figure*}
\figurebox{45pc}{}{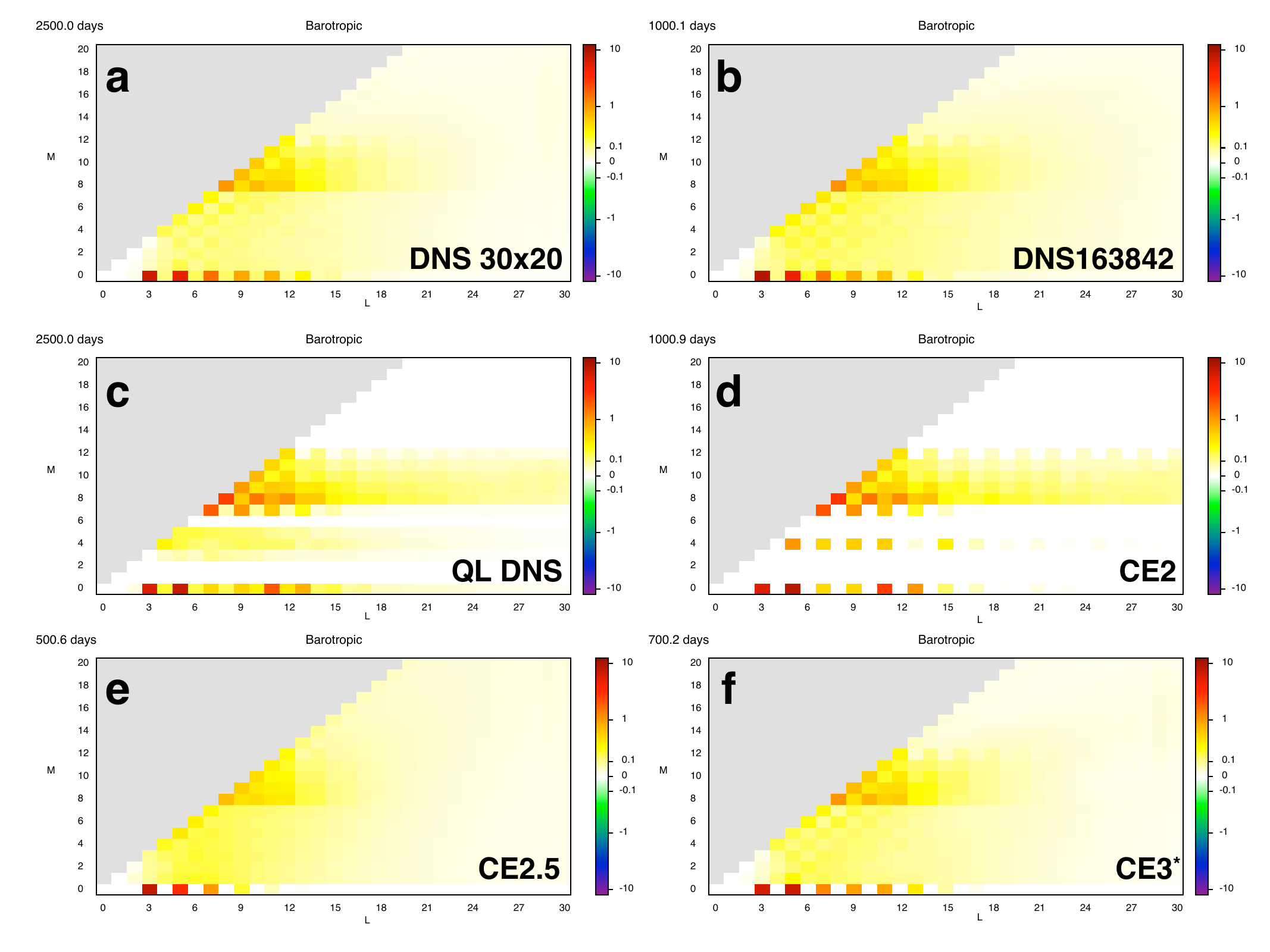}
\caption{Spectral power of relative vorticity in each mode as calculated by DNS and DSS.  Vertical axis: Zonal wavenumber $m$.  Horizontal axis: Spherical wavenumber $\ell$.  As in Figure \ref{timelines}.}
\label{power}
\end{figure*}

The power spectrum of the relative vorticity field is shown in Figure \ref{power}.  As expected, CE2 and QL DNS do not pass power scale-by-scale from the forcing scales to either longer or shorter scales (there are no cascades) as is made evident by the clean gaps in spectral power.  By contrast, CE2.5 and CE3$^*$ distribute power throughout spectral space, similar to DNS, showing that cascades, while not dominant, are still operating.     

\begin{figure*}
\figurebox{50pc}{}{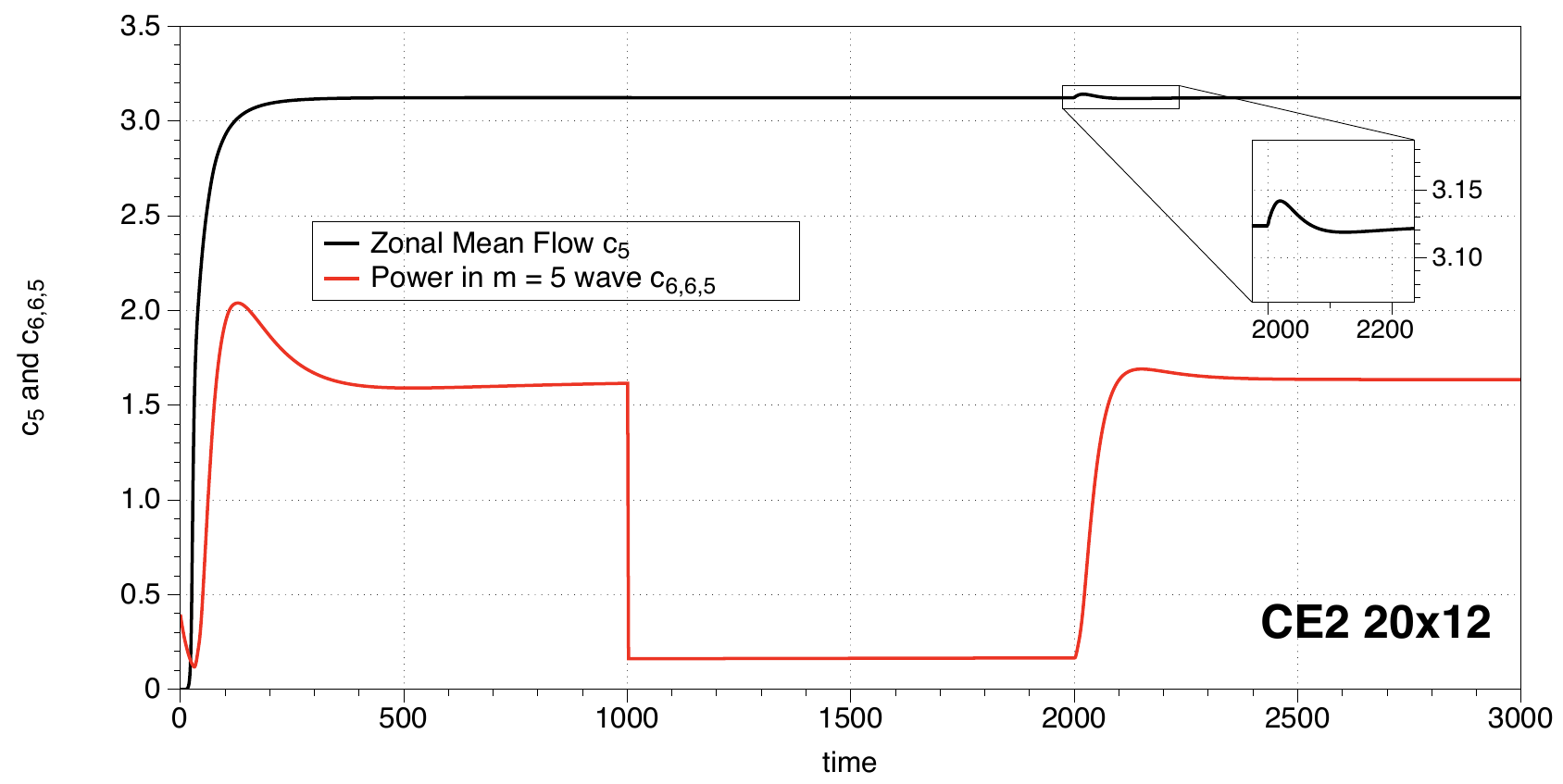}
\caption{CE2 simulation experiment with $L = 20$ and $M = 12$ demonstrates the existence of a Rossby wave that is generated as a result of an instability in the mean flow of the jet.   The wave has zonal wavevector $m = 5$ which is one of the wave vectors that is not forced stochastically.   At time $t = 1,000$ the first cumulant is frozen, and power in the $m = 5$ wave is decreased by a factor of $10$ (the figure shows only the $\ell = 6$ component of the wave; all components behave the same).  At a later time $t = 2,000$ the hold on the first cumulant is released, allowing it to evolve again towards the fixed point.  The first cumulant makes an adjustment that causes the wave to become temporarily unstable, grow, and finally saturate back at its fixed point value.}
\label{experiment}
\end{figure*}

Interestingly both CE2 and QL DNS develop a separate mode at $m = 4$ (CE2) and over $3 \leq m \leq 5$ (QL DNS).  (North-south reflection symmetry across the equator is imposed upon CE2.  Turning it off leads to spectra identical to QL DNS.)    As these zonal modes are not stochastically driven, the Rossby waves that form instead have their origin in an instability that appears as the zonal mean flow becomes established.   The physics is similar to the instability of a flow driven towards a prescribed unstable jet \citep{marstonconoveretal2008} as can be demonstrated by a numerical experiment on CE2.\footnote{We thank C. Nardini for suggesting the experiment.}   Figure \ref{experiment} shows the result (at a reduced resolution $L = 20$ and $M = 12$ to suppress the slow oscillations in the zonal means).  At time $t = 1,000$ CE2 has reached the fixed point, time evolution of the first cumulant is stopped, and power in the $m = 5$ wave is decreased by reducing the $m = 5$ part of the second cumulant by a factor of $10$ (the figure shows only the $\ell = 6$ component of the wave).  The $m = 5$ mode remains neutrally stable, neither growing nor dissipating despite drag acting upon it.  At a later time $t = 2,000$ the hold on the first cumulant is released, allowing the CE2 system to again evolve towards the fixed point.  The first cumulant makes an adjustment that temporarily causes the wave to become unstable, grow, and then saturate again back to its fixed point value.  

A small peak in spectral power appears even in DNS at zonal wavenumber $m = 5$ (not shown), demonstrating that though CE2 exaggerates its strength, it does capture important physics.  That such waves can appear in CE2 appears to have been missed in much of the literature.   It would be interesting to investigate the role of the waves further, including a possible connection to oscillating zonal means.  

\begin{figure*}
\figurebox{45pc}{}{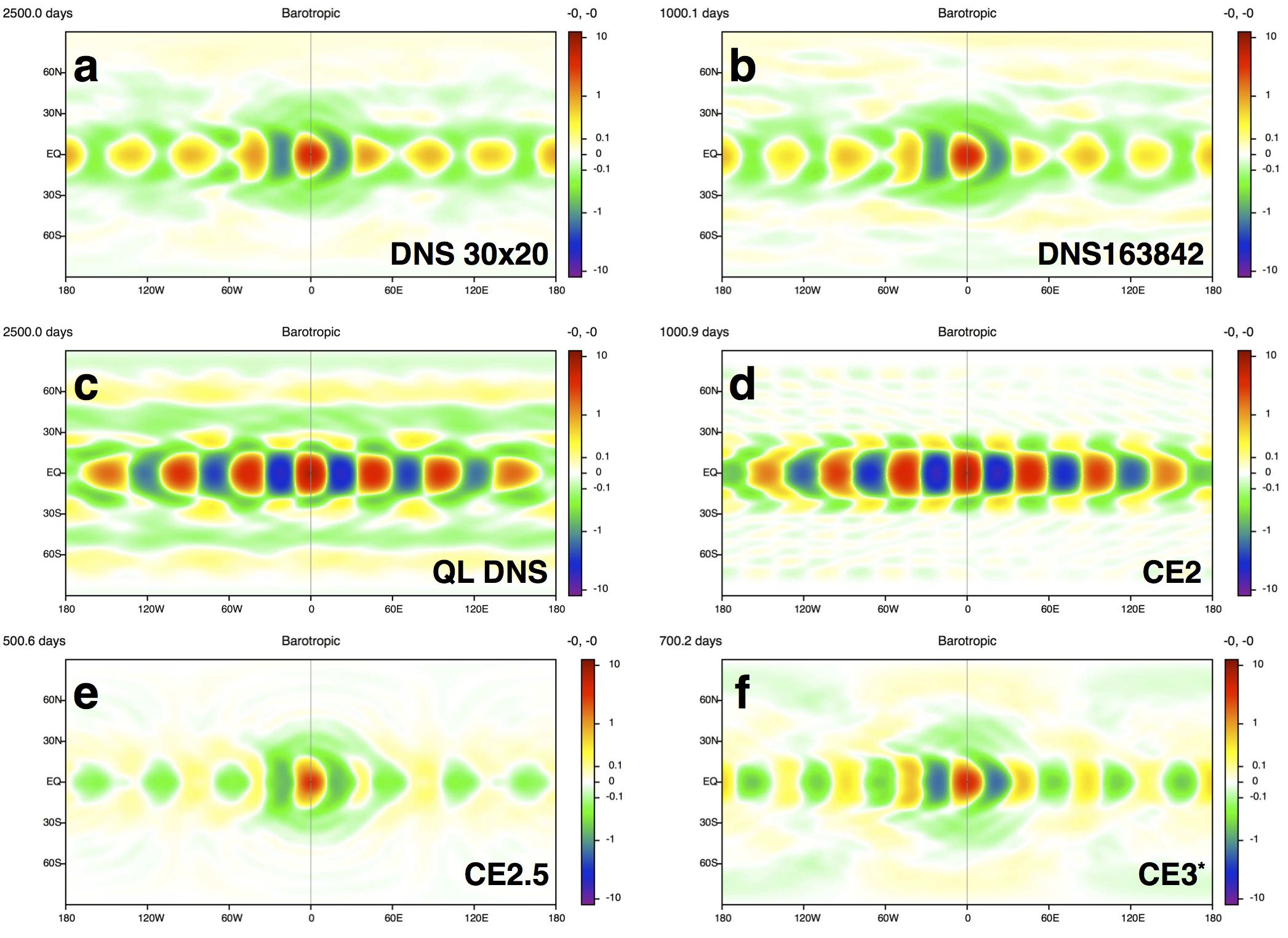}
\caption{Second cumulant (two-point correlation function of the vorticity).  One point is centered along the prime meridian at latitude $0^\circ$.  The non-local nature of the correlations or teleconnections is evident.  As in Figure \ref{timelines}.}
\label{twoPoint0}
\end{figure*}

\begin{figure*}
\figurebox{45pc}{}{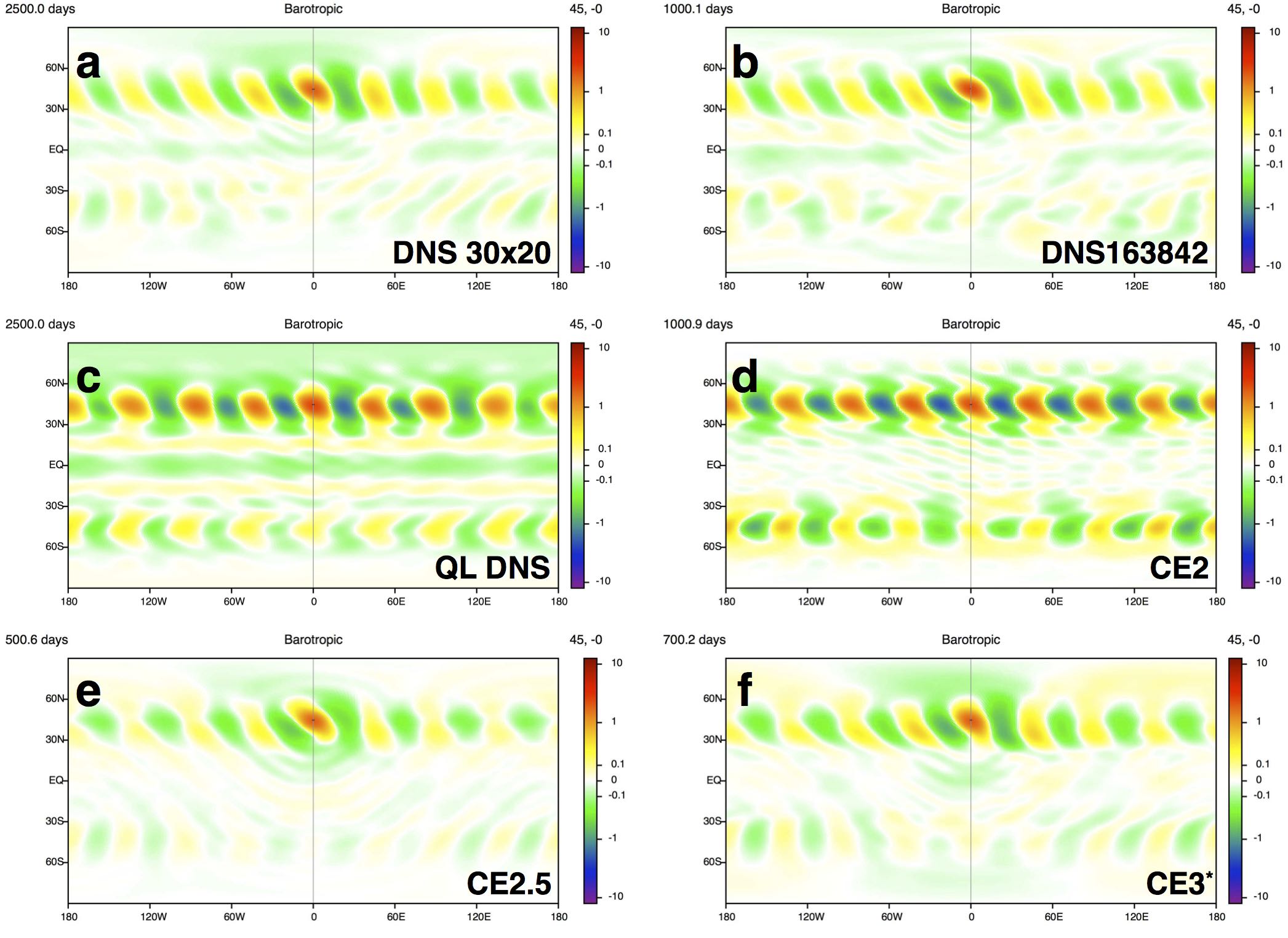}
\caption{Second cumulant (two-point correlation function of the vorticity).  One point is centered along the prime meridian at latitude $45^\circ$.  
The anisotropy and inhomogeneity of the statistic is plain.  As in Figure \ref{timelines}.}
\label{twoPoint45}
\end{figure*}

\begin{figure*}
\figurebox{25pc}{}{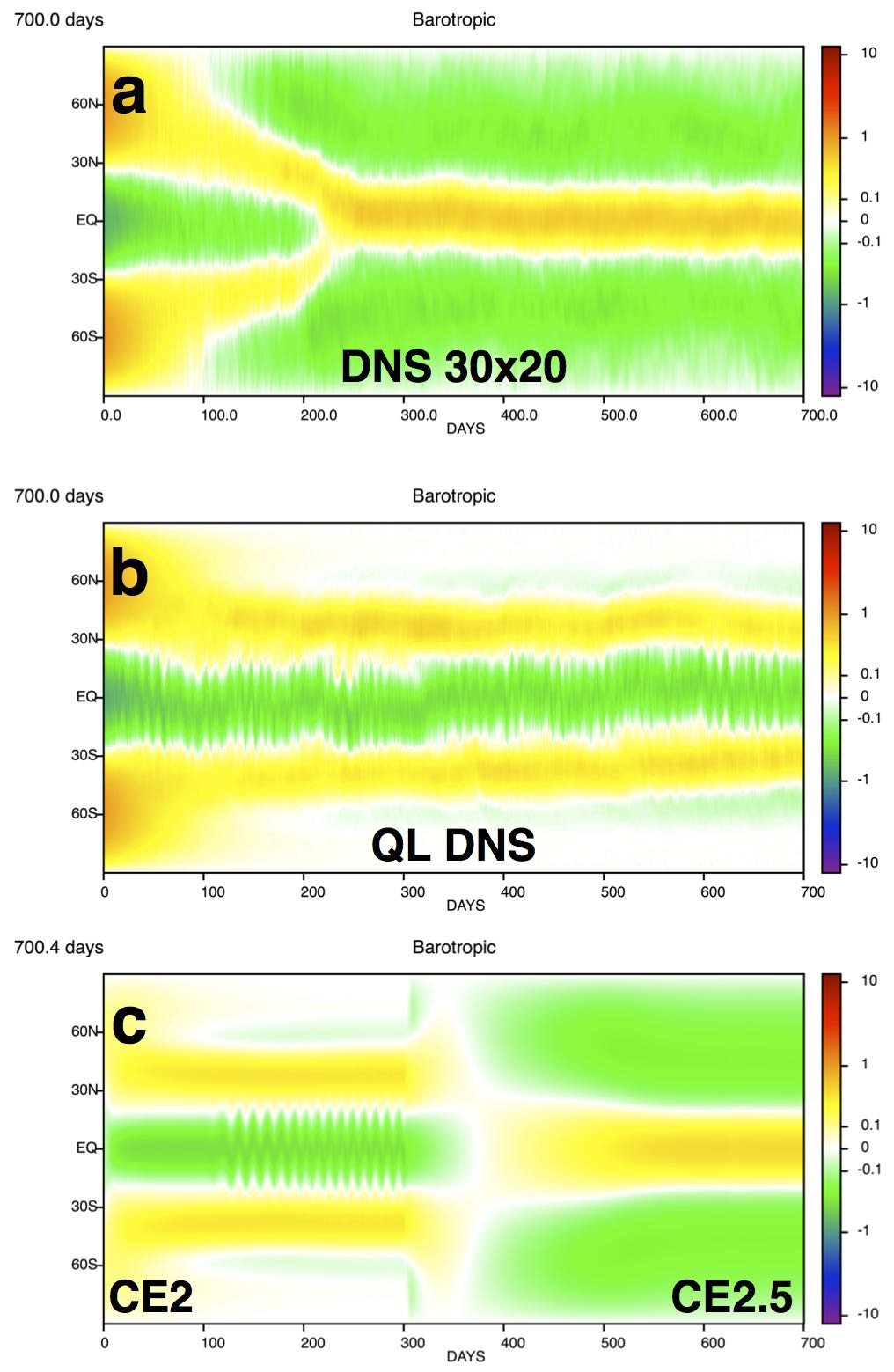}
\caption{Hovm\"oller timeline of the zonal mean zonal velocity for a jet with initial reversal of direction.  (a) Spectral DNS shows that the equatorial jet reverts to a prograde flow after a couple hundred days.  In a quasi-linear simulation, the equatorial jet remains retrograde and does not return to a prograde direction, even out to $t = 1200$. (c) At time $t = 0$ the first cumulant of the CE2 calculation is initialized such that a retrograde equatorial jet forms.  North-south reflection symmetry about the equator is not imposed and the zonal mean oscillates much as it does in the quasi-linear simulation. The equatorial jet is trapped in a prograde direction.  At time $t = 300$ the CE2.5  (with $\tau = 2$) correction is turned on, and the jet reverses, matching those found in DNS.}
\label{reversal}
\end{figure*}

The non-local, anisotropic, and inhomogeneous nature of correlations is especially manifest in plots of the two-point correlation function (the second cumulant) shown in Figures \ref{twoPoint0} and \ref{twoPoint45}.  Here the exaggerated coherent waves of CE2 and QL DNS are also evident.  The higher-order closures do a much better job of reproducing the incoherence seen in DNS.   Another weakness of CE2 (and QL DNS) is apparent in Figure \ref{reversal}.  Here the jet problem is initialized with a strong retrograde equatorial jet.  Full DNS reverses the jets after a couple hundred days, restoring the prograde orientation of the equatorial jet.  QL DNS and CE2 are both locked in the incorrect configuration, even when north-south reflection symmetry is no longer imposed.  Longer simulations show that they remain stuck.  Turning on the CE2.5 correction to CE2, however, restores the jets to their correct orientation.  

\subsection{Summary}
 
Comparison between DNS and the several types of DSS examined in this Chapter shows that DSS can accurately capture the low-order equal-time statistics of the stochastically-driven barotropic jet.  Qualitative agreement found at the CE2 level is systematically improved by the inclusion of higher-order corrections (CE2.5 and CE3$^*$).  CE2 is fast -- faster even than DNS -- but suffers from a lack of predictive power:  Different initializations lead to different stationary statistics, including reversed jets and even absence of north-south reflection symmetry.  (QL DNS suffers from the same weaknesses).  This is one manifestation of the existence of multiple equilibria \citep{parkerkrommes2013a,parkerkrommes2014,constantinoufarrelletal2014}.  The higher-order closures correct these defects, and can even restore north-south reflection symmetry (when it is not already imposed).  CE2.5 and CE3$^*$ are similar, suggesting that accurately representing the third cumulant is less important than the contribution that it makes to the second cumulant.  However, the higher-order methods come at the cost of much higher computational effort.

\section{Conclusions}
\label{Conclusions}

We conclude this Chapter by summarizing our results and discussing the pros and cons of DSS.  We have described how the problems of the formation and maintenance of jets naturally lend themselves to solution via this technique. Owing to their importance in geophysical and astrophysical fluid dynamics --- as evidenced by the contributions in the rest of this Volume --- the jet problem may act as a benchmark for theory, whether statistical in origin or not. Furthermore, the non-trivial nature of the interactions that lead to zonal flow formation (which is sometimes characterized as the turbulence acting as a negative viscosity) are replicated in many other systems of geophysical and astrophysical interest.

We stress however that the simplest level of approximation for DSS (CE2) does not always yield results that accurately represent the statistics obtained from DNS. Simply put, for systems far from statistical equilibrium higher-level approximations are needed to reproduce the covariances and even the qualitative behavior seen by accumulating statistics obtained from DNS.  We have shown how to include systematically corrections at the CE3$^*$ or CE2.5 levels that appear to be realizable and (in the case of CE2.5) conserve global invariants. Furthermore these systems {\em do} appear to yield statistics that compare well with those obtained from DNS. 
Solving the equations at these higher levels of truncation, however, does represent a considerable computational challenge. For this reason we are currently researching whether a generalization of CE2 that includes multiple zonal modes at the largest scales, and not just the $m = 0$ mode, is capable of yielding accurate representations at reduced computational cost. 

Recent work is extending DSS well beyond single-layer barotropic problems.  Deterministic baroclinic models of planetary atmospheres (including primitive equations) are being studied \citep{ogormanschneider2007,marston2010,marston2012,aitchaalalschneider2014,aitchaalalmarstonetal2014}.  A model of a magnetized stellar tachocline has been studied \citep{tobiasdagonetal2011}.  Turbulence generated in three dimensions by shear can be examined 
\citep{constantinounavidetal2014}.  Also a first application of CE2 to three-dimensional magnetohydrodynamics has now appeared: See \citet{squirebhattacharjee2014}.  

We conclude by stating that we believe that DSS forms an important complement to other approaches, such as Direct Numerical Simulation, analytic closure theories, weak/wave turbulence theory and of course laboratory experiments; all of which are discussed elsewhere in this book as applied to the problem of the formation and maintenance of jets. Direct Statistical Simulation provides valuable insights into the important physical processes in a given problem, by highlighting the key interactions and cross-correlations. Moreover because DSS relies on systematic approximations, it is ideal for isolating the root causes of emerging phenomena. This in turn should lead to the construction of better models for these phenomena with all the critical processes included. 
Perhaps more importantly, DSS gives the potential for modeling physical systems in more extreme parameter regimes. Astrophysical and geophysical flows are often found in parameter ranges extremely far from the region of applicability of theory and for which direct computation is prohibitively expensive even with efficient codes optimized for massively parallel architectures. Because DSS solves directly for the statistics of the flows, which are smoother in space and have less complicated temporal behavior, the solution takes the form of evolution on a simple manifold that may be accessed using extremely efficient algorithms. For this reason we believe that DSS could prove essential in elucidating the behavior of geophysical and astrophysical flows in parameter regimes that will remain inaccessible to theory and DNS for many years to come.

\begin{acknowledgment}
\chapter*{Acknowledgments}
We are grateful for helpful discussions with Farid Ait-Chaalal, Freddy Bouchet, Greg Chini, James Cho, Brian Farrell, Boris Galperin, Petros Ioannou, John Krommes, Cesare Nardini, Jeff Parker, Peter Read, Tapio Schneider, Tomas Tangarife, and Bill Young.  We thank the Isaac Newton Institute for Mathematical Sciences where work on this Chapter was initiated during the Program on ``Mathematics for the Fluid Earth.''   JBM also would like to thank Freddy Bouchet, ENS-Lyon, and CNRS for hosting a visit there when this work was completed.  This work was also supported in part by the NSF under grant Nos. DMR-0605619 and CCF-1048701 (JBM and WQ) and DMR-1306806 (JBM). 
\end{acknowledgment}

\index{Direct Statistical Simulation}


\begin{thebibliography}{66}
\expandafter\ifx\csname natexlab\endcsname\relax\def\natexlab#1{#1}\fi
\expandafter\ifx\csname selectlanguage\endcsname\relax
  \def\selectlanguage#1{\relax}\fi

\bibitem[\protect\citename{Ait-Chaalal and Schneider,
  }2014]{aitchaalalschneider2014}
Ait-Chaalal, Farid, and Schneider, Tapio. 2014.
\newblock {Why eddy momentum fluxes are concentrated in the upper troposphere
  }.
\newblock {\em Journal of the Atmospheric Sciences}, {\bf 72}, 1585--1064.

\bibitem[\protect\citename{Ait~Chaalal {et~al.},
  }2016]{aitchaalalmarstonetal2014}
Ait~Chaalal, Farid, Schneider, Tapio, Meyer, Bettina, and Marston, J.~B. 2016.
\newblock Cumulant expansions for atmospheric flows.
\newblock {\em New Journal of Physics}, {\bf 18}, 025019.

\bibitem[\protect\citename{Arakawa, }1966]{arakawa1966}
Arakawa, Akio. 1966.
\newblock {Computational design for long-term numerical integration of the
  equations of fluid motion: Two-dimensional incompressible flow. Part I}.
\newblock {\em J. Comp. Phys.}, {\bf 1}, 119--143.

\bibitem[\protect\citename{Bakas and Ioannou, }2013a]{bakasioannou2013}
Bakas, N.~A., and Ioannou, P.~J. 2013a.
\newblock On the mechanism underlying the spontaneous emergence of barotropic
  zonal jets.
\newblock {\em J. Atmos.\ Sci.}, {\bf 70}, 2251--2271.

\bibitem[\protect\citename{Bakas and Ioannou, }2014]{bakasioannou2013d}
Bakas, Nikolaos, and Ioannou, Petros. 2014.
\newblock A theory for the emergence of coherent structures in beta-plane
  turbulence.
\newblock {\em Journal of Fluid Mechanics}, {\bf 740}, 312--341.

\bibitem[\protect\citename{Bakas and Ioannou, }2011]{bakasioannou2011}
Bakas, Nikolaos~A., and Ioannou, Petros~J. 2011.
\newblock Structural stability theory of two-dimensional fluid flow under
  stochastic forcing.
\newblock {\em J. Fluid Mech.}, {\bf 682}, 332--361.

\bibitem[\protect\citename{Bakas and Ioannou, }2013b]{bakasioannou2013b}
Bakas, Nikolaos~A., and Ioannou, Petros~J. 2013b.
\newblock Emergence of Large Scale Structure in Barotropic $\beta$-Plane
  Turbulence.
\newblock {\em Phys. Rev. Lett.}, {\bf 110}, 224501.

\bibitem[\protect\citename{Bakas and Ioannou, }2013c]{bakasioannou2013c}
Bakas, Nikolaos~A., and Ioannou, Petros~J. 2013c.
\newblock A theory for the emergence of coherent structures in beta-plane
  turbulence.
\newblock {\em Journal of Fluid Mechanics}, {\bf 740}, 312 -- 341.

\bibitem[\protect\citename{Bartello and Holloway, }1991]{bartelloholloway1991}
Bartello, Peter, and Holloway, Greg. 1991.
\newblock Passive scalar transport in $\beta$-plane turbulence.
\newblock {\em Journal of Fluid Mechanics}, {\bf 223}(2), 521--536.

\bibitem[\protect\citename{Bertoglio, }2003]{bertoglio2003}
Bertoglio, Jean-Pierre. 2003.
\newblock {Two-point closures and turbulence modeling}.
\newblock {In:} {\em {3rd Int. Symp. on Turbulence and Shear Flow Phenomena}}.

\bibitem[\protect\citename{Bouchet {et~al.}, }2013]{bouchetnardinietal2013}
Bouchet, Freddy, Nardini, Cesare, and Tangarife, Tom\'as. 2013.
\newblock Kinetic Theory of Jet Dynamics in the Stochastic Barotropic and 2D
  Navier-Stokes Equations.
\newblock {\em Journal of Statistical Physics}, {\bf 153}(4), 572--625.

\bibitem[\protect\citename{Bowman and Krommes, }1997]{bowmankrommes1997}
Bowman, J.~C., and Krommes, J.~A. 1997.
\newblock The realizable {M}arkovian closure and realizable test-field model.
  {II}: {A}pplication to anisotropic drift-wave turbulence.
\newblock {\em Phys.\ Plasmas}, {\bf 4}, 3895--3909.

\bibitem[\protect\citename{Bowman {et~al.}, }1993]{bowmankrommesetal1993}
Bowman, J.~C., Krommes, J.~A., and Ottaviani, M. 1993.
\newblock The realizable {M}arkovian closure. {I}. {G}eneral theory, with
  application to three-wave dynamics.
\newblock {\em Phys. Fluids B}, {\bf 5}, 3558--3589.

\bibitem[\protect\citename{{Canuto} and {Minotti}, }2001]{canutominotti2001}
{Canuto}, V.~M., and {Minotti}, F. 2001.
\newblock {Mixing and transport in stars - I. Formalism: momentum, heat and
  mean molecular weight}.
\newblock {\em Mon. Not. Roy. Ast. Soc.}, {\bf 328}, 829--838.

\bibitem[\protect\citename{Constantinou {et~al.},
  }2014a]{constantinoufarrelletal2014}
Constantinou, Navid~C, Farrell, Brian~F, and Ioannou, Petros~J. 2014a.
\newblock {Emergence and equilibration of jets in beta-plane turbulence:
  applications of Stochastic Structural Stability Theory}.
\newblock {\em Journal of the Atmospheric Sciences}, {\bf 71}, 1818 -- 1842.

\bibitem[\protect\citename{Constantinou {et~al.},
  }2014b]{constantinounavidetal2014}
Constantinou, Navid~C, Lozano-Dur{\'a}n, Adrian, Nikolaidis, Marios-Andreas,
  Farrell, Brian~F, Ioannou, Petros~J, and Jim{\'e}nez, Javier. 2014b.
\newblock {Turbulence in the highly restricted dynamics of a closure at second
  order: comparison with DNS}.
\newblock {\em Journal of Physics: Conference Series}, {\bf 506}, 012004.

\bibitem[\protect\citename{Davidson, }2004]{davidson2004}
Davidson, P.~A. 2004.
\newblock {\em Turbulence: An Introduction for Scientists and Engineers}.
\newblock Oxford University Press.

\bibitem[\protect\citename{DelSole, }2001]{delsole2001}
DelSole, T. 2001.
\newblock A theory for the forcing and dissipation in stochastic turbulence
  models.
\newblock {\em J. Atmos.\ Sci.}, {\bf 58}, 3762--3775.

\bibitem[\protect\citename{Domaradzki and Orszag, }1987]{domaradzkietal1987}
Domaradzki, J~Andrzej, and Orszag, Steven~A. 1987.
\newblock {Numerical solutions of the direct interaction approximation
  equations for anisotropic turbulence}.
\newblock {\em Journal of Scientific Computing (ISSN 0885-7474)}, {\bf 2},
  227--248.

\bibitem[\protect\citename{Farrell and Ioannou, }2009]{farrellioannou2009}
Farrell, B.~F., and Ioannou, P.~J. 2009.
\newblock A theory of baroclinic turbulence.
\newblock {\em J. Atmos.\ Sci}, {\bf 66}, 2444--2454.

\bibitem[\protect\citename{Farrell and Ioannou, }2007]{farrellioannou2007}
Farrell, Brian~F., and Ioannou, Petros~J. 2007.
\newblock Structure and Spacing of Jets in Barotropic Turbulence.
\newblock {\em J. Atmos. Sci.}, {\bf 64}, 3652--3665.

\bibitem[\protect\citename{Frederiksen, }1999]{frederiksen1999}
Frederiksen, J.~S. 1999.
\newblock Subgrid-scale parameterizations of eddy-topographic force, eddy
  viscosity, and stochastic backscatter for flow over topography.
\newblock {\em J. Atmos.\ Sci.}, {\bf 56}, 1481--1494.

\bibitem[\protect\citename{Frederiksen, }2012]{frederiksen2012b}
Frederiksen, Jorgen~S. 2012.
\newblock {Self-Energy Closure for Inhomogeneous Turbulent Flows and Subgrid
  Modeling}.
\newblock {\em Entropy}, {\bf 14}(4), 769--799.

\bibitem[\protect\citename{Frisch, }1995]{Frisch1995}
Frisch, U. 1995.
\newblock {\em Turbulence: The Legacy of A. N. Kolmogorov}.
\newblock Cambridge: Cambridge University Press.

\bibitem[\protect\citename{Galperin {et~al.},
  }2006]{galperinsukorianskyetal2006}
Galperin, B., Sukoriansky, S., Dikovskaya, N., Read, P.~L., Yamazaki, Y.~H.,
  and Wordsworth, R. 2006.
\newblock Anisotropic turbulence and zonal jets in rotating flows with a
  $\beta$ effect.
\newblock {\em Nonlinear Processes Geophys.}, {\bf 13}, 83--98.

\bibitem[\protect\citename{Galperin {et~al.},
  }2010]{galperinsukorianskyetal2010}
Galperin, Boris, Sukoriansky, Semion, and Dikovskaya, Nadejda. 2010.
\newblock Geophysical flows with anisotropic turbulence and dispersive waves:
  flows with a $\beta$-effect.
\newblock {\em Ocean Dynamics}, {\bf 60}(2), 427--441.

\bibitem[\protect\citename{Heikes and Randall, }1995a]{heikesrandall1995a}
Heikes, R., and Randall, D.~A. 1995a.
\newblock {Numerical integration of the shallow-water equations on a twisted
  icosahedral grid. Part I. Basic design and results of tests.}
\newblock {\em Mon.~Wea.~Rev.}, {\bf 123}, 1862--1880.

\bibitem[\protect\citename{Heikes and Randall, }1995b]{heikesrandall1995b}
Heikes, R., and Randall, D.~A. 1995b.
\newblock {Numerical integration of the shallow-water equations on a twisted
  icosahedral grid. Part II. A detailed description of the grid and an analysis
  of numerical accuracy.}
\newblock {\em Mon.~Wea.~Rev.}, {\bf 123}, 1881--1887.

\bibitem[\protect\citename{Herr {et~al.}, }1996]{herretal1996}
Herr, Stacy, Wang, Lian-Ping, and Collins, Lance~R. 1996.
\newblock {EDQNM model of a passive scalar with a uniform mean gradient}.
\newblock {\em Physics of Fluids}, {\bf 8}(6), 1588--1608.

\bibitem[\protect\citename{Herring, }1963]{herring1963}
Herring, J~R. 1963.
\newblock {Investigation of problems in thermal convection.}
\newblock {\em Journal of Atmospheric Sciences}, {\bf 20}(4), 325--338.

\bibitem[\protect\citename{Kraichnan, }1980]{kraichnan1980}
Kraichnan, R~H. 1980.
\newblock {Realizability inequalities and closed moment equations}.
\newblock {\em International Conference on Nonlinear Dynamics}, {\bf 357},
  37--46.

\bibitem[\protect\citename{{Krause} and {Raedler}, }1980]{krauseraedler1980}
{Krause}, F., and {Raedler}, K.~H. 1980.
\newblock {\em {Mean-field magnetohydrodynamics and dynamo theory}}.
\newblock Pergamon Press.

\bibitem[\protect\citename{Legras, }1980]{legras1980}
Legras, Bernard. 1980.
\newblock {Turbulent phase shift of rossby waves}.
\newblock {\em Geophysical {\&} Astrophysical Fluid Dynamics}, {\bf 15}(1),
  253--281.

\bibitem[\protect\citename{Lilly, }1969]{lilly1969}
Lilly, Douglas~K. 1969.
\newblock Numerical simulation of two-dimensional turbulence.
\newblock {\em Physics of Fluids}, {\bf 12}, II--240--249.

\bibitem[\protect\citename{Maltrud and Vallis, }1991]{maltrudvallis1991}
Maltrud, M~E, and Vallis, G~K. 1991.
\newblock {Energy spectra and coherent structures in forced two-dimensional and
  beta-plane turbulence}.
\newblock {\em Journal of Fluid Mechanics}, {\bf 228}, 321--342.

\bibitem[\protect\citename{{Marston}, }2010]{marston2010}
{Marston}, J.~B. 2010.
\newblock {Statistics of the general circulation from cumulant expansions}.
\newblock {\em Chaos}, {\bf 20}(4), 041107.

\bibitem[\protect\citename{Marston, }2012]{marston2012}
Marston, J.~B. 2012.
\newblock Planetary atmospheres as nonequilibrium condensed matter.
\newblock {\em Annu.\ Rev.\ Condensed Matter Phys.}, {\bf 3}, 285--310.

\bibitem[\protect\citename{{Marston} {et~al.}, }2008]{marstonconoveretal2008}
{Marston}, J.~B., {Conover}, E., and {Schneider}, T. 2008.
\newblock {Statistics of an Unstable Barotropic Jet from a Cumulant Expansion}.
\newblock {\em Journal of Atmospheric Sciences}, {\bf 65}, 1955.

\bibitem[\protect\citename{O'Gorman and Schneider, }2007]{ogormanschneider2007}
O'Gorman, PA, and Schneider, T. 2007.
\newblock {Recovery of atmospheric flow statistics in a general circulation
  model without nonlinear eddy--eddy interactions}.
\newblock {\em Geophys. Res. Lett}, {\bf 34}, 524--535.

\bibitem[\protect\citename{O'Kane and Frederiksen, }2004]{okanefrederiksen2004}
O'Kane, Terrence~J, and Frederiksen, Jorgen~S. 2004.
\newblock {The QDIA and regularized QDIA closures for inhomogeneous turbulence
  over topography}.
\newblock {\em J. Fluid Mech.}, {\bf 504}, 133--165.

\bibitem[\protect\citename{Orszag, }1970]{orszag1970}
Orszag, S.~A. 1970.
\newblock Anaytical theories of turbulence.
\newblock {\em J. Fluid Mech.}, {\bf 41}, 363--386.

\bibitem[\protect\citename{Orszag, }1977]{orszag1977}
Orszag, Steven~A. 1977.
\newblock {Lectures on the Statistical Theory of Turbulence}.
\newblock {\em Fluid Dyanmics, Les Houches 1973}.

\bibitem[\protect\citename{Parker and Krommes, }2013]{parkerkrommes2013a}
Parker, Jeffrey~B., and Krommes, John~A. 2013.
\newblock Zonal flow as pattern formation.
\newblock {\em Physics of Plasmas}, {\bf 20}(10), 100703.

\bibitem[\protect\citename{Parker and Krommes, }2014]{parkerkrommes2014}
Parker, Jeffrey~B, and Krommes, John~A. 2014.
\newblock {Generation of zonal flows through symmetry breaking of statistical
  homogeneity}.
\newblock {\em New Journal of Physics}, {\bf 16}, 035006.

\bibitem[\protect\citename{Qi and Marston, }2014]{qimarston2014}
Qi, W, and Marston, J~B. 2014.
\newblock {Hyperviscosity and statistical equilibria of Euler turbulence on the
  torus and the sphere}.
\newblock {\em Journal of Statistical Mechanics: Theory and Experiment}, {\bf
  2014}(7), P07020.

\bibitem[\protect\citename{Qi, }2014]{qi2014}
Qi, Wanming. 2014.
\newblock {\em {Statistical Approaches to Two-Dimensional Turbulence.}}
\newblock Ph.D. thesis, Brown University.

\bibitem[\protect\citename{Read {et~al.}, }2007]{readyamazakietal2007}
Read, Peter~L., Yamazaki, Y.~H., Lewis, S.~R., Williams, P.~D., Wordsworth, R.,
  Miki-Yamazaki, K., Sommeria, J., and Didelle, H. 2007.
\newblock Dynamics of Convectively Driven Banded Jets in the Laboratory.
\newblock {\em J. Atmos. Sci.}, {\bf 64}, 4031--4052.

\bibitem[\protect\citename{Rhines, }1975]{rhines1975}
Rhines, P.~B. 1975.
\newblock Waves and turbulence on a beta-plane.
\newblock {\em J. Fluid Mech.}, {\bf 69}, 417--443.

\bibitem[\protect\citename{Rhines, }1979]{rhines1979}
Rhines, P.~B. 1979.
\newblock Geostrophic turbulence.
\newblock {\em Annu.\ Rev.\ Fluid Mech.}, {\bf 11}, 401--441.

\bibitem[\protect\citename{{Ruediger}, }1989]{ruediger1988}
{Ruediger}, G. 1989.
\newblock {\em {Differential rotation and stellar convection. Sun and the solar
  stars}}.
\newblock Akademie Verlag.

\bibitem[\protect\citename{Saad, }2003]{saad2003}
Saad, Y. 2003.
\newblock {\em Iterative methods for sparse linear systems}.
\newblock Society for Industrial Mathematics.

\bibitem[\protect\citename{{Salmon}, }1998]{salmon1998}
{Salmon}, R. 1998.
\newblock {\em {Lectures on Geophysical Fluid Dynamics}}.
\newblock Oxford University Press.

\bibitem[\protect\citename{Sapsis and Majda, }2013a]{SapsisMajda2013a}
Sapsis, T~P, and Majda, A~J. 2013a.
\newblock {Statistically accurate low-order models for uncertainty
  quantification in turbulent dynamical systems}.
\newblock {\em Proceedings of the National Academy of Sciences}, {\bf 110}(34),
  13705 -- 13710.

\bibitem[\protect\citename{Sapsis and Majda, }2013b]{SapsisMajda2013b}
Sapsis, Themistoklis~P, and Majda, Andrew~J. 2013b.
\newblock {A statistically accurate modified quasilinear Gaussian closure for
  uncertainty quantification in turbulent dynamical systems}.
\newblock {\em Physica D}, {\bf 252}, 34--45.

\bibitem[\protect\citename{{Scott} and {Polvani}, }2008]{scottpolvani2008}
{Scott}, R.~K., and {Polvani}, L.~M. 2008.
\newblock {Equatorial superrotation in shallow atmospheres}.
\newblock {\em Geophys. Res. Lett.}, {\bf 35}, 24202.

\bibitem[\protect\citename{Scott and Dritschel, }2012]{scottdritschel2012}
Scott, Richard~K., and Dritschel, David~G. 2012.
\newblock The structure of zonal jets in geostrophic turbulence.
\newblock {\em J. Fluid Mech.}, {\bf 711}, 576--598.

\bibitem[\protect\citename{Silberman, }1954]{silberman1954}
Silberman, Isadore. 1954.
\newblock {Planetary Waves in the Atmosphere.}
\newblock {\em Journal of Atmospheric Sciences}, {\bf 11}(1), 27--34.

\bibitem[\protect\citename{Squire and Bhattacharjee,
  }2015]{squirebhattacharjee2014}
Squire, Jonathan, and Bhattacharjee, Amitava. 2015.
\newblock {Statistical simulation of the magnetorotational dynamo}.
\newblock {\em Physical Review Letters}, {\bf 114}, 085002.

\bibitem[\protect\citename{Srinivasan and Young, }2012]{srinivasanyoung2012}
Srinivasan, K., and Young, W.~R. 2012.
\newblock Zonostrophic instability.
\newblock {\em J. Atmos.\ Sci.}, {\bf 69}, 1633--1656.

\bibitem[\protect\citename{Thiebaux, }1971]{thiebaux1971}
Thiebaux, ML. 1971.
\newblock {On the Structure of Interaction Coefficients in the Spectral
  Equations for Planetary Waves.}
\newblock {\em Journal of Atmospheric Sciences}, {\bf 28}, 1294--1294.

\bibitem[\protect\citename{Tobias and Marston, }2013]{tobiasmarston2013}
Tobias, S.~M., and Marston, J.~B. 2013.
\newblock Direct Statistical Simulation of Out-of-Equilibrium Jets.
\newblock {\em Phys. Rev. Lett.}, {\bf 110}, 104502.

\bibitem[\protect\citename{Tobias {et~al.}, }2011]{tobiasdagonetal2011}
Tobias, S.~M., Dagon, K., and Marston, J.~B. 2011.
\newblock Astrophysical fluid dynamics via direct statistical simulation.
\newblock {\em Astrophys.~J.}, {\bf 727}, 127--138.

\bibitem[\protect\citename{Vallis and Maltrud, }1993]{vallismaltrud1993}
Vallis, G.~K., and Maltrud, M.~E. 1993.
\newblock Generation of mean flows and jets on a beta-plane and over
  topography.
\newblock {\em J. Phys. Oceanog.}, {\bf 23}, 1346--1362.

\bibitem[\protect\citename{{Vallis}, }2006]{vallis2006}
{Vallis}, Geoffrey.~K. 2006.
\newblock {\em {Atmospheric and Oceanic Fluid Dynamics}}.
\newblock Cambridge University Press.

\bibitem[\protect\citename{Warneford and Dellar, }2013]{warneforddellar2013}
Warneford, Emma~S, and Dellar, Paul~J. 2013.
\newblock {The quasi-geostrophic theory of the thermal shallow water
  equations}.
\newblock {\em J. Fluid Mech.}, {\bf 723}, 374--403.

\bibitem[\protect\citename{Williams, }2009]{williams2009}
Williams, P.D. 2009.
\newblock {A Proposed Modification to the Robert--Asselin Time Filter}.
\newblock {\em Monthly Weather Review}, {\bf 137}, 2538--2546.

\end{thebibliography}
\end{document}